\documentclass[review]{elsarticle}
\usepackage{lineno,hyperref}
\modulolinenumbers[5]
\usepackage[utf8]{inputenc}
\usepackage[francais,english]{babel}		                 

\usepackage{amsmath}						 
\usepackage{amssymb}						 
\usepackage[makeroom]{cancel}

\usepackage{graphicx}
\usepackage{dcolumn}
\usepackage{bm}
\usepackage{multirow}
\usepackage{textcomp}
\usepackage[]{subfigure}
\usepackage{booktabs}
\usepackage{array}
\usepackage{pgf}
\usepackage{graphicx}
\usepackage{bm}
\usepackage[toc,page]{appendix}
\usepackage{hyperref}

\usepackage{tikz}

\begin{document}
\begin{frontmatter}
%
\title{Modelling the relationship between deformed microstructures and static recrystallization textures: application to ferritic stainless steels}


\author[ubc,INP]{A.~Després\corref{cor1}}
	\ead{arthur.despres@grenoble-inp.fr}
\author[APERAM]{J.~D.~Mithieux}
\author[ubc]{C.~W.~Sinclair}

\address[ubc]{Department of Materials Engineering, The University of British Columbia, 309-6350 Stores Road, Vancouver, Canada} 
\address[INP]{Univ. Grenoble Alpes, CNRS, Grenoble INP, SIMaP, F-38000, Grenoble, France}
\address[APERAM]{APERAM, F-62330, Isbergues, France}

\begin{abstract}

We present an original approach for predicting the static recrystallization texture development during annealing of deformed crystalline materials. The microstructure is considered as a population of subgrains and grains whose sizes and boundary properties determine their growth rates. The model input parameters are measured directly on orientation maps maps of the deformed microstructure measured by electron backscattered diffraction. The anisotropy in subgrain properties then drives a competitive growth giving rise to the recrystallization texture development. The method is illustrated by a simulation of the static recrystallization texture development in a hot rolled ferritic stainless steel. The model predictions are found to be in good agreement with the experimental measurements, and allow for an in-depth investigation of the formation sequence of the recrystallization texture. A distinction is established between the texture components which develop due to favorable growth conditions and those developing due to their predominance in the prior deformed state. The high fraction of $\alpha$ fibre orientations in the recrystallized state is shown to be a consequence of their predominance in the deformed microstructure rather than a preferred growth mechanism. A close control of the fraction of these orientations before annealing is thus required to minimize their presence in the recrystallized state.

\end{abstract}


\end{frontmatter}

\section{Introduction}

The crystallographic texture development occuring during static recrystallization of polycrystalline materials is commonly thought to result from a complex combination of microstructural features in the deformed state. In first approximation, orientations whose grains are associated with boundaries of high energy and mobility are most likely to develop \cite{rollett_growth_1997,humphreys_unified_1997-1,brechet_nucleation_2006}. The anisotropy and spatial distribution of stored energy and subgrain size may also play a role in the texture development. In materials deformed to moderate strains and at high temperatures, recrystallized grains tend to have orientations associated with low stored energy, while in materials deformed to high strains and at cold temperatures, high stored energy orientations are usually prefered \cite{humphreys_recrystallization_2017,doherty_current_1997}.

The two principal approaches to express the relationship between deformed microstructures and recrystallization textures regard the recrystallization texture development either from the perspective of `nucleation' or `growth', although some models consider both aspects. In models focused on growth, the orientation relationship between the deformed microstructure and the orientations known to recrystallize is investigated. For example, the model of Bunge and Köhler investigates the orientations most likely to grow in the deformed textures of fcc and bcc metals \cite{bunge_model_1992}. Engler implemented a similar approach to investigate orientation pinning, i.e. the selective slowing down of specific recrystallized grains with the progress of recrystallization \cite{engler_influence_1998}. In the model of Sebald and Gottstein, the mobility of the recrystallized grains depends on the orientation relationship between the recrystallized and deformed grain orientations \cite{sebald_modeling_2002}. The general conclusion of these phenomenological models is that the texture components known to develop in recrystallized grains must be, from a statistical point of view, in favourable conditions for growing in a microstructure whose orientations correspond to the deformed texture.

As the recrystallization texture is a product of the deformed microstructure, several attempts have also been made to link the preferential nucleation of orientations to their behaviour during plastic deformation. Differences in nucleation rates between the texture components, have been, for example, attributed to intergranular slip activity contrasts (estimated by Taylor factors) \cite{kestens_modeling_1996,sidor_modeling_2011}, resolved shear stresses \cite{wenk_deformation-based_1997,lebensohn_modelling_1998}, and intragranular disorientation levels \cite{zecevic_modelling_2019} calculated from crystal plasticity simulations. In a recent work, Steiner \emph{et al.} \cite{steiner_monte_2017} related the preferential nucleation of orientations to measurements of kernel average disorientation made on electron backscattered diffraction maps of the deformed microstructure.

As pointed out by Raabe \cite{raabe_23_2014}, the inhomogeneities that lead to the formation of recrystallized grains are often overlooked to maintain computational efficiency of models. Thus, in most cases, the success in predicting the texture results more from a careful choice of rules and simulation parameters than from an apropriate description of the mechanisms driving recrystallization. These models are, of course, helpful to identify the first-order parameters leading the texture development, but their reliability and their sensitivity to the state of the microstructure is limited in proportion to these assumptions.

In this article, we present a cellular growth model for predicting static recrystallization textures which takes as input the experimentally measured characteristics of deformed microstructures. The model is adapted from an earlier theoretical work \cite{despres_mean-field_2020}, where it was assumed that recrystallization occurs by the competitive growth of subgrains in the deformed microstructure. The initial subgrain properties, which drive the recrystallization kinetics and the development of texture, are calculated directly from orientation maps measured by electron backscattered diffraction. 

The model is illustrated by simulating the recrystallization texture developed during annealing of a hot rolled ferritic stainless steel sheet. This is an interesting test case for several reasons. Ferritic stainless steels have a high stacking-fault energy and readily form subgrains during deformation, in particular hot deformation. In addition, recrystallization of hot rolled ferritic stainless steels is known to develop so called $\alpha$ fibre orientations which are deleterious for formability and ridging \cite{zhang_effects_2011,raabe_textures_1993,mehtonen_microstructural_2014}, and which demand significant efforts to be removed during the subsequent processing steps. Little information is currently available in the literature about the origin of this texture, and, importantly, it is not reproduced by simulations.

In what follows, experimental measurements of the microstructure and texture evolution during recrystallization of the selected steel are presented first. Next, the model is described. This is followed by its application to the experimentally investigated case. A discussion is finally conducted on the origin of the recrystallization texture of hot rolled ferritic stainless steels and on the aspects of processing which it may be sensitive to.

\section{The experimental data}

\subsection{Materials and processing}

The experiments were conducted using a AISI445 grade of ferritic stainless steel grade produced by APERAM. The nominal composition of the alloy is given in \autoref{tab:Compo}. The bcc phase is stable at all temperatures up to the melting point. Titanium and niobium may precipitate into carbides and nitrides, but at the temperatures investigated these particles are large and present in a sufficiently small fraction that their effect on recrystallization can be neglected \cite{jacquet_etude_2013}. 

\begin{table}[htbp]
\centering
\caption{Nominal composition of the grade studied, in wt.\%}
\begin{tabular}{ c c c c c c c c  }
  C & N & Si & Mn & Cr & Ti+Nb  & Cu & Fe \\
  \hline
  0.015 & 0.028 & 0.25 & 0.25 & 20.20 & 0.65 &  0.45 & bal.\\
  
\end{tabular}
\label{tab:Compo}
\end{table}

A 100mm long, 80mm wide and 15mm thick strip was machined from the center of an industrially produced transfer bar (i.e. the product between the roughing and finishing mills in hot rolling). This strip was reheated in a box furnace at 1150$^\circ$C during 40mn, before being rolled using a laboratory rolling mill. The sample was air-cooled to 1100$^\circ$C, before being rolled to 75\% thickness reduction in one pass and then water quenched. The rolling temperature was monitored with thermocouples inserted at the center of the thickness. The average strain rate during rolling was estimated to 23s$^{-1}$, while the Zener-Hollomon parameter was $\sim$3$\times$10$^{13}$s$^{-1}$ (see ref. \cite{despres_origin_2019} for more details). The lab scale rolling operation was performed at the OCAS R$\&$D center in Ghent, Belgium.

Next, samples of 60mm$\times$10mm$\times$3.75mm were prepared from the middle of the width of the hot rolled strip, with the length parallel to the rolling direction. These were annealed in a DSI Gleeble thermomechanical apparatus to induce static recrystallization of the deformed microstructures. The samples were annealed at 1100$^\circ$C for respectively 2s, 5s and 12s, with a preliminary heating step of 3s. Temperature was monitored and controlled by thermocouples placed at the center of the samples. At the end of the treatment, the samples were water-quenched.

\subsection{Observation tools}

Microstructures and textures of the deformed and recrystallized samples were characterized from orientation maps obtained by electron back-scattered diffraction (EBSD). Samples were prepared by standard mechanical polishing down to 1$\mu$m diamond surface finish, followed by electropolishing in a solution of 5\% perchloric acid + 95\% acetic acid. Observations were conducted at the location of the Gleeble thermocouples, in the central third of the sample thickness, where the deformation results from approximately plane strain compression \cite{engler_study_2000,roumina_deformation_2008}. 
All maps were acquired with a step size of 1$\mu$m to allow for differentiation of subgrains and recrystallized grains at the same time. After acquisition, maps were pre-processed using a Kuwahara filter to reveal subgrains separated by boundaries with disorientation angle of 0.3$^\circ$ or more (i.e. the standard angular resolution obtained by this method \cite{brough_optimising_2006,humphreys_orientation_2001}). Details about the implementation of the Kuwahara filter are given in \autoref{app:kuwa}. While small regions of interest will be shown in the illustrative figures below, it is important to have in mind that the results were obtained for each condition from several large maps to ensure representativity of the measurements (as a rule of thumb, 5000 to 10000 individual orientations are required for proper texture calculation \cite{wright_comparison_2007}). Finally, orientation distribution functions (ODF) were calculated with the Kernel method of MTEX, with a half-width of 5$^\circ$ and assuming orthotropic sample symmetry. ODFs were plotted in the $\varphi_2=45^\circ$ section of Euler space as it contains most of the important texture information for ferritic stainless steels \cite{engler_study_2000}.

On the filtered orientation maps, subgrains were defined as regions enclosed by boundaries of disorientation angle higher than 0.3$^\circ$ and whose equivalent-area diameter is smaller than recrystallized grains. By contrast, recrystallized grains were defined as regions enclosed by boundaries of disorientation angle higher than 0.3$^\circ$ whose equivalent-area diameter is above 30$\mu$m. Note that, as for any recrystallization study, the definition of recrystallized grains is motivated by practical considerations and only aims to provide a threshold that describes correctly the observations. No requirement was imposed on the presence of high-angle boundaries surrounding recrystallized grains as these are often but not systematically in contact with high-angle boundaries \cite{wright_review_2011}.


%



\subsection{Measurements of recrystallization}

\begin{figure}[htbp]
\centering

\subfigure[]{\includegraphics[width=0.32\textwidth]{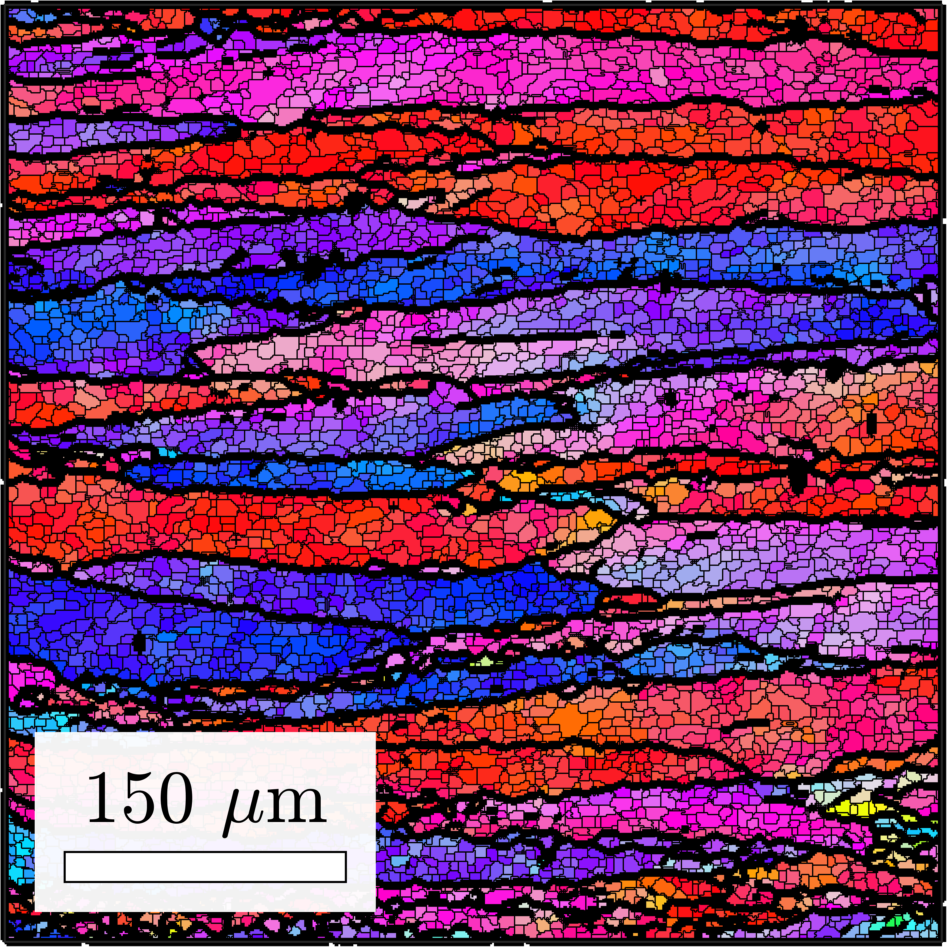}}
\subfigure[]{\includegraphics[width=0.32\textwidth]{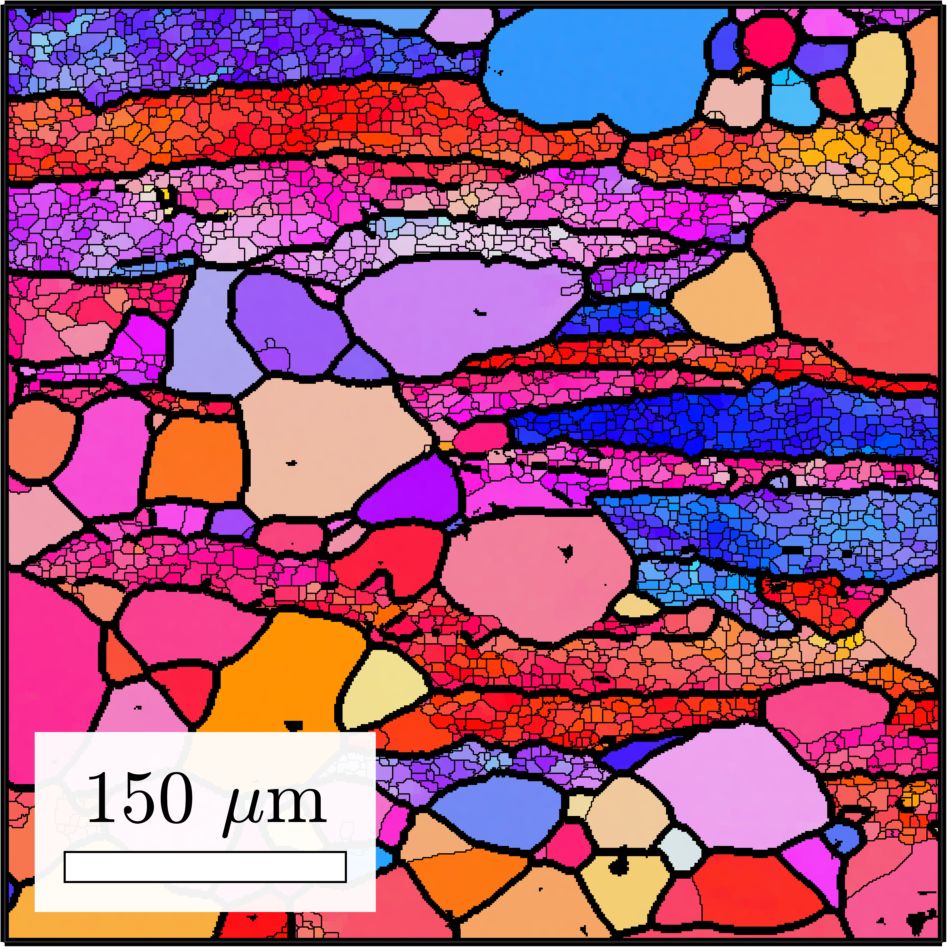}}
\subfigure[]{\includegraphics[width=0.32\textwidth]{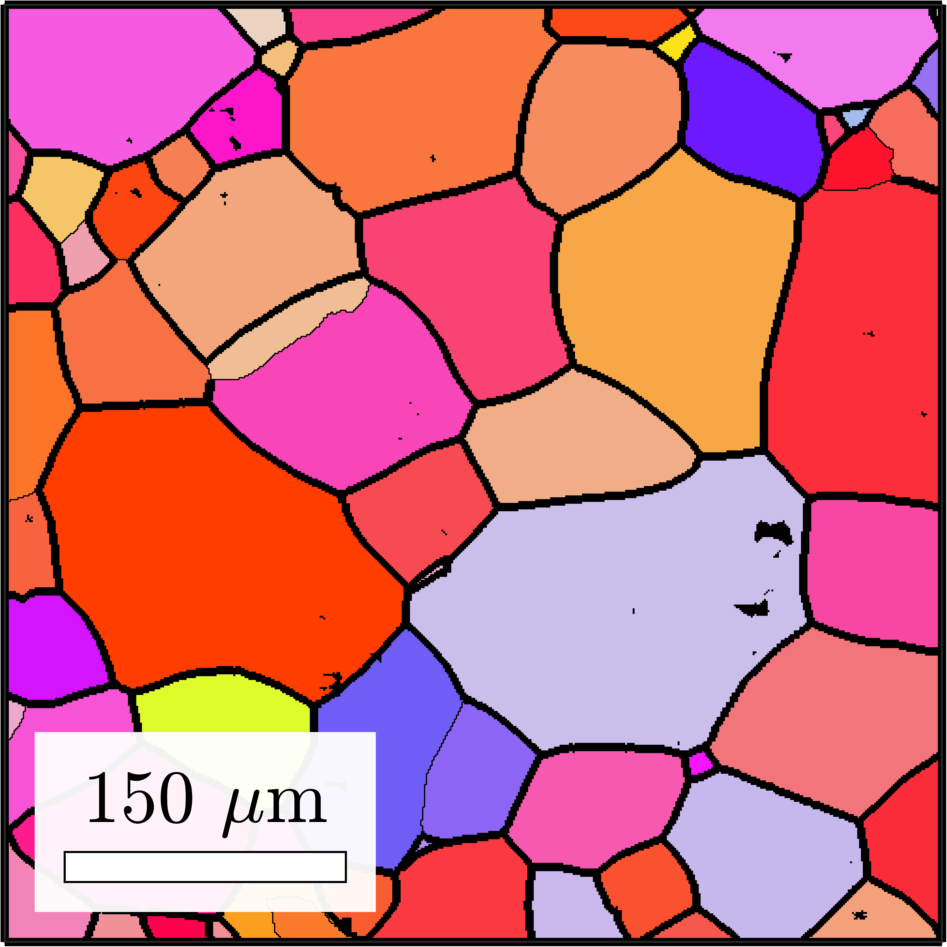}}

\subfigure[]{\includegraphics[width=0.32\textwidth]{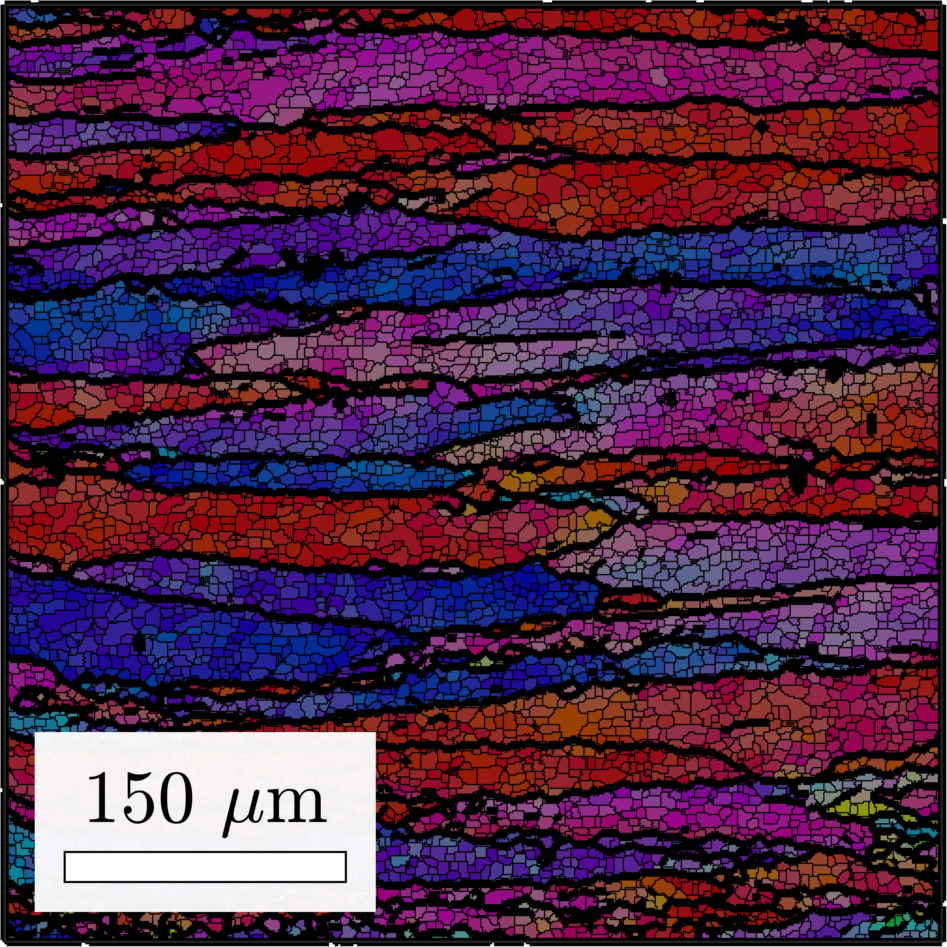}}
\subfigure[]{\includegraphics[width=0.32\textwidth]{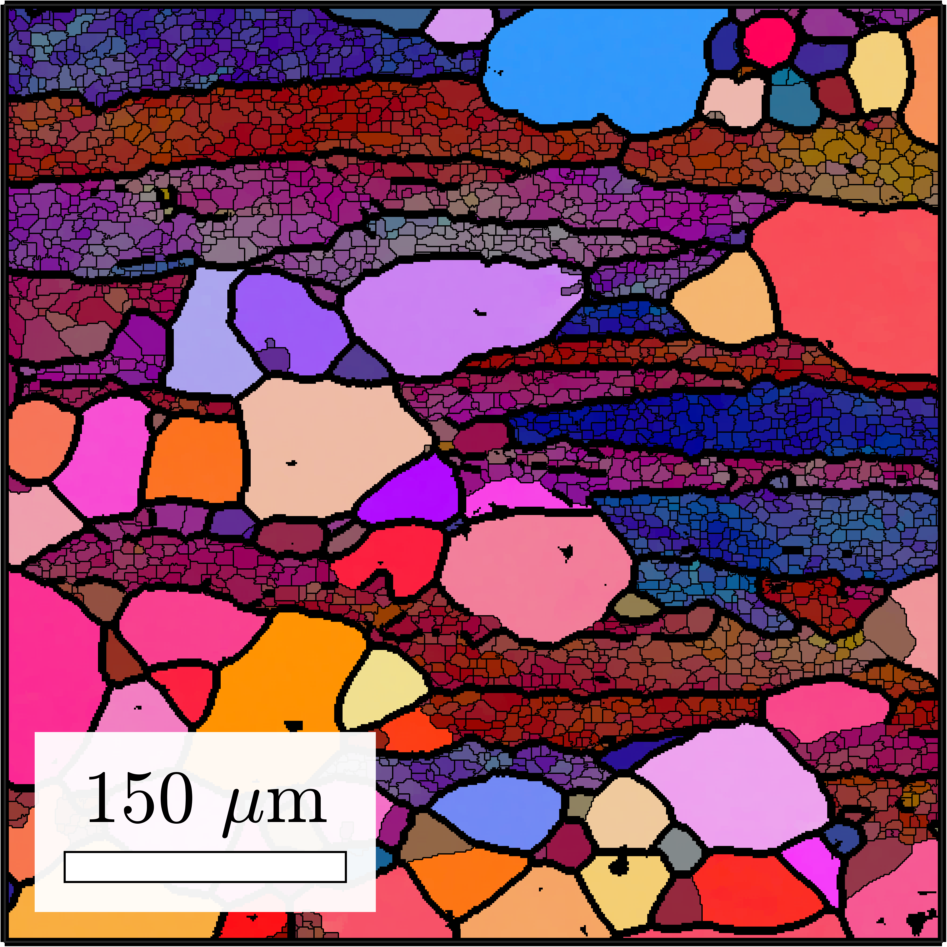}}
\subfigure[]{\includegraphics[width=0.32\textwidth]{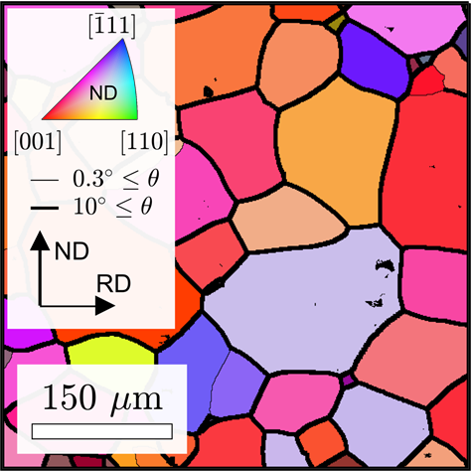}}
\caption{ND-IPF maps of the microstructure a) after deformation, b) after 2s annealing at 1100$^\circ$C, c) after 12s annealing at 1100$^\circ$C. In d-e), the identified recrystallized grains are highlighted with enhanced brightness. The maps have been pre-processed with the Kuwahara filter.}
\label{fig:ebsd-maps}
\end{figure}

\autoref{fig:ebsd-maps} shows inverse pole figure (IPF) maps of the microstructure after deformation and annealing. The colours represent the crystallographic axis parallel to the sheet's normal direction (ND). The microstructure after deformation (\autoref{fig:ebsd-maps}a) is composed of deformed grains elongated in the rolling direction (RD). These appear as large agregates of subgrains, are surrounded mostly by boundaries of misorientation angle higher than 10$^\circ$, and exhibit rather homogeneous internal orientation spreads. Their interior is composed of numerous subgrains of mean equivalent-area diameter equal to 6.85$\mu$m. Note that about 1.8$\times$10$^5$ subgrains were measured for this condition. The magnitude of the subgrain size, i.e. a few microns, is consistent with other work on hot deformed ferritic stainless steels \cite{despres_origin_2019,mehtonen_dynamic_2014}. As no subgrain exceeds the threshold recrystallized grain diameter (\autoref{fig:ebsd-maps}d), the recrystallized fraction is 0\% in this state.

After 2s of annealing (\autoref{fig:ebsd-maps}b), the microstructure has coarsened in a heterogeneous way, giving rise to a few subgrains much larger than others. The subgrains above the threshold recrystallized grain diameter are identified as recrystallized grains (\autoref{fig:ebsd-maps}e), and the calculated recrystallized fraction is 44\%.  The recrystallized grain boundaries are preferentially of high-angle, but do not exhibit particular orientation relationships. For example, the fraction of boundaries having a disorientation angle above 10$^\circ$ is 75\% between recrystallized grains and non-recrystallized regions of the microstructure, while it is 94\% if taken between recrystallized grains themselves. Assuming a 5$^\circ$ deviation around the exact relationship, boundaries with the $\Sigma19$a relationship (i.e. 27$^\circ$ rotation around the $\langle110\rangle$ axis), make up only 1.3\% of boundaries between recrystallized grains and deformed grains, and 1.4\% of those between recrystallized grains themselves. These boundaries have been suggested to be very important for recrystallization textures in ferritic stainless steels \cite{raabe_selective_1992}. 



After 12s of annealing, the initial microstructure has been fully replaced by large grains surrounded mostly by high angle boundaries (\autoref{fig:ebsd-maps}c). The calculated recrystallized fraction is 96\%. In essence, the microstructure at this stage is fully recrystallized. Indeed, with the chosen definition of recrystallized grains, the recrystallized fraction cannot reach exactly 100\% since there will always be grains whose apparent diameter on an image in two dimension is below the threshold recrystallized grain diameter (\autoref{fig:ebsd-maps}f). The fraction of boundaries with disorientation angles above 10$^\circ$ is 94\% for the whole microstructure. It is also noticeable that the recrystallized grain diameters have become much larger than the initial deformed grain size in the normal direction. This indicates that recrystallized grains grow mostly towards the deformed grains neighbouring their parent deformed grains. As, in addition, most high-angle boundaries in the deformed microstructure are pre-existing deformed grain boundaries, one may conclude that recrystallization occurs by the bulging of deformed grain boundaries. This interpretation is consistent with the opinion that materials deformed at high temperature and low strain rate recrystallize by this mechanism \cite{humphreys_recrystallization_2017,bate_re-evaluation_1997}.

\begin{figure}[htbp]
\centering
\subfigure[]{\includegraphics[width=0.32\textwidth]{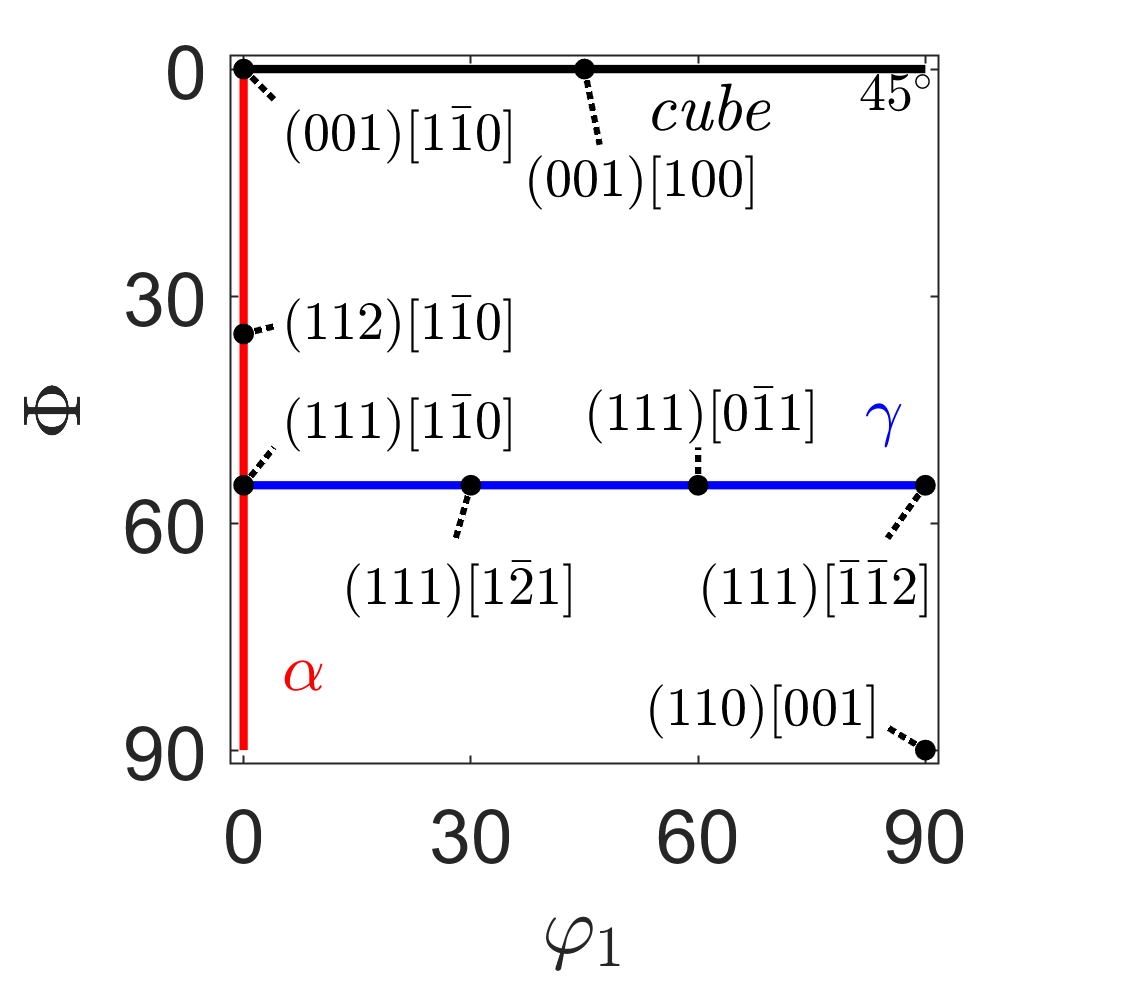}}
\subfigure[]{\includegraphics[width=0.32\textwidth]{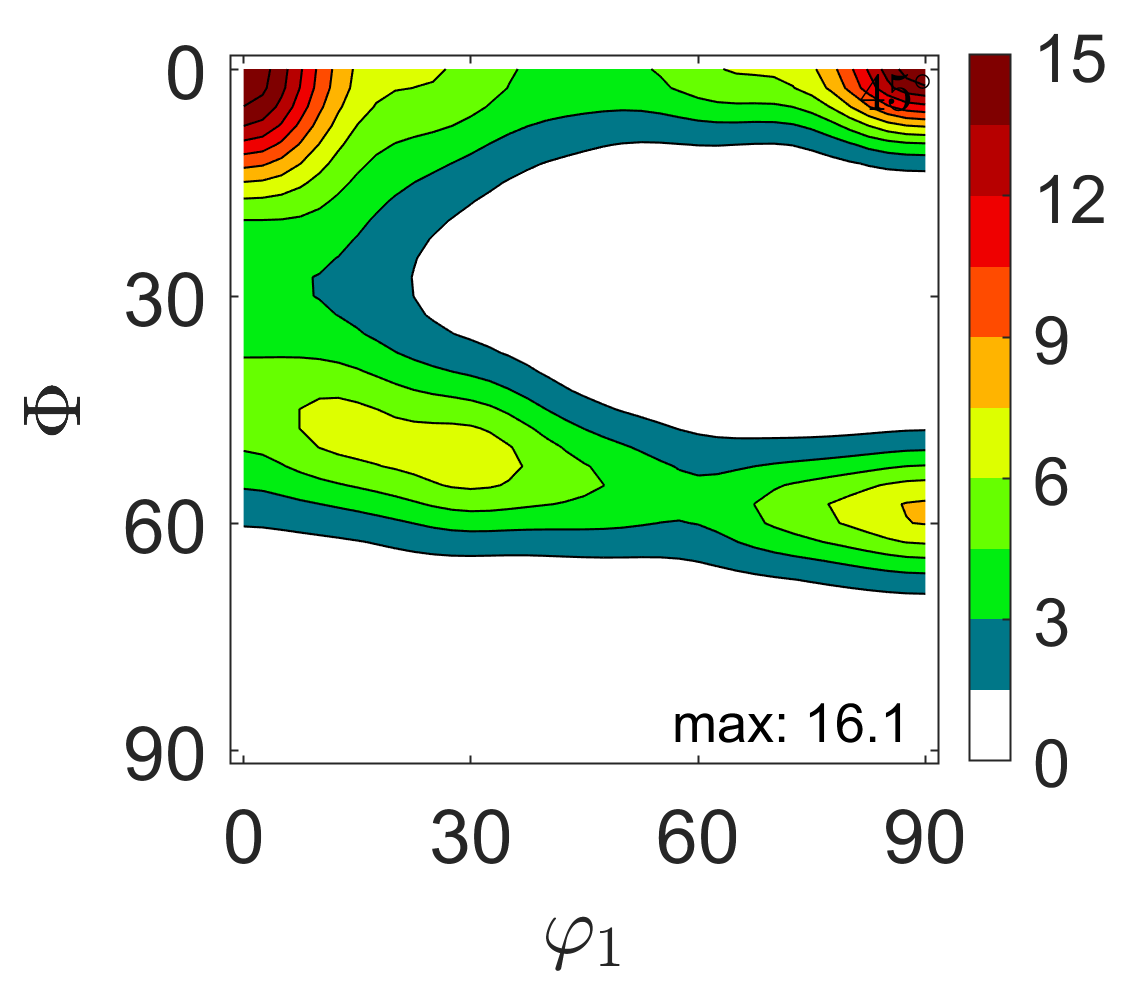}}
\subfigure[]{\includegraphics[width=0.32\textwidth]{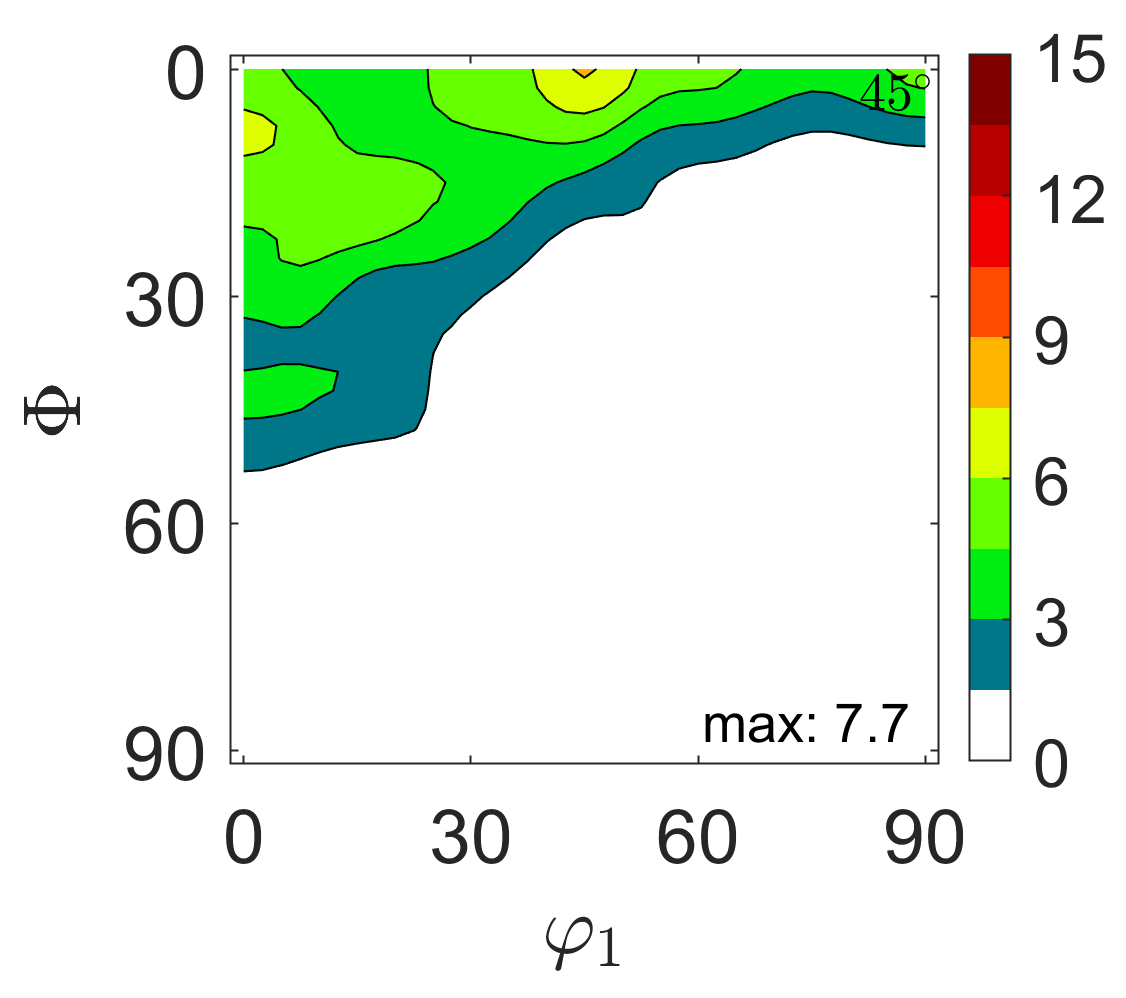}}
\caption{a) position of the low-index components in the $\varphi_2=45^\circ$ section of the Euler space. Note the distinction between the cube orientation $\{001\}\langle100\rangle$ and the cube fibre orientation $\{001\}\langle uvw\rangle$. b) ODF of the deformed microstructure, c) ODF of the microstructure after 12s of annealing. The scalebar is in multiple of random (m.r.d).}
\label{fig:odf-exp}
\end{figure}

\autoref{fig:odf-exp} shows the effect static recrystallization on texture. To help the reader, the location in Euler space of several low-index components is shown in \autoref{fig:odf-exp}a. The deformation texture shown in \autoref{fig:odf-exp}b is composed of strong $\alpha$, $\gamma$ and cube fibre orientations, with a maximum intensity at the rotated cube orientation $\{001\}\langle110\rangle$. This texture is common to hot rolled sheets of many ferritic steels \cite{zhang_effects_2011,mehtonen_microstructural_2014,jacquet_etude_2013,hutchinson_deformation_1999}. Following the classic argument \cite{kocks_texture_2000}, the exceptional strength of the rotated cube orientation is likely caused by dynamic recrystallization occuring during deformation as it cannot be reproduced by crystal plasticity simulations \cite{despres_origin_2019}. 

After 12s of annealing (full recrystallization), the texture has weakened significantly. The cube fibre and upper portion of the $\alpha$ fibre remain the preferred orientations, with the two strongest components located near the rotated cube orientation $\{001\}\langle110\rangle$ and at the cube orientation $\{001\}\langle100\rangle$. Again, the ODF is very similar to that of previous reports on recrystallization of hot rolled ferritic steels \cite{jacquet_etude_2013,jonas_effects_2001,mehtonen_microstructural_2014}.

To complement this analysis, \autoref{tab:volume-fractions} provides a measure of the texture evolution in terms of volume fraction of low-index components in the deformation and recrystallization texture. The values confirm the weakening of the rotated cube orientation $\{001\}\langle110\rangle$ and the strengthening of the cube orientation $\{001\}\langle100\rangle$. Interestingly, even if the Goss orientation $\{110\}\langle001\rangle$ is a minor component in both textures, it strenghens proportionally more than any of the major components. This aspect will be discussed later in this article.

\begin{table}[htbp]
\centering
\caption{Volume fraction of orientations within $\pm$15$^\circ$ of the low index components in the deformed and 12s annealed (i.e. recrystallized) samples.}
\begin{tabular}{ c c c c c c c  }
   & $\{001\}\langle110\rangle$ & $\{112\}\langle110\rangle$ & $\{111\}\langle110\rangle$ & $\{111\}\langle121\rangle$ & $\{001\}\langle100\rangle$ & $\{110\}\langle001\rangle$ \\
  \hline
  Def. & 18.1\% & 6.2\% & 5.6\% & 8.4\% & 5.53\% & 0.1\%\\
   Rex. & 8.1\% & 5.3\% & 2.9\% & 1.8\% & 10.2\% & 2.1\%\\
  
\end{tabular}
\label{tab:volume-fractions}
\end{table}

\section{Model}

The modelling approach presented in this article relies extensively on assumptions developed in an earlier work \cite{despres_mean-field_2020}, where recrystallization was simulated by cellular growth laws applied to a population of subgrains pre-existing in the deformed microstructure. This assumption is common for high stacking-fault energy materials \cite{jacquet_etude_2013,hurley_modelling_2003,zurob_quantitative_2006,ferry_discontinuous_1996},  and seems apropriate for the case under investigation. 

A key difference must be noted between the work in ref. \cite{despres_mean-field_2020} and the present work. In \cite{despres_mean-field_2020}, the subgrain boundary properties were estimated statistically for a synthetic microstructure exhibiting intragranular recrystallization, i.e. for a case where recrystallized grains grow within their parent deformed grain. In the present work, recrystallization is intergranular since the recrystallized grains grow towards neighbouring deformed grains. In this configuration, recrystallization depends mostly on the subgrains in contact with the deformed grain boundaries. This feature cannot be accounted for by the statistical approach developed previously. Therefore, in the following sections, a different approach is implemented to extract the boundary properties directly from the experimental measurements.



\subsection{Growth laws}

The microstructure is considered as a set of individual spherical grains and subgrains characterized by diameter $D_{(i)}$, mean boundary energy $\Gamma_{(i)}$ and mean boundary mobility $M_{(i)}$, embedded in a homogeneous medium of properties $\bar{D}$ and $\bar{\Gamma}$. The model makes no distinction between grains and subgrains, the same laws being applied to all objects. Assuming that these parameters are known, the growth rate of a grain or subgrain in three dimension is given by \cite{despres_mean-field_2020}:

%

\begin{equation}
\frac{dD_{\left(i\right)}}{dt}=\frac{2M_{\left(i\right)}\Gamma_{\left(i\right)}}{D_{\left(i,t\right)}}\left(a_{\left(i\right)}\left(3+\frac{16D_{\left(i,t\right)}}{9\bar{D}_{(t)}}\right)-5\right)
\label{eq:growth-rate3}
\end{equation}

Where the capillary term $a_{(i)}= 6 sin^{-1}(\Gamma_{(i)}/ 2\bar{\Gamma})/\pi \leq 3$ accounts for the variations of boundary curvature as a function of the heterogeneity of boundary energy between the individual grain or subgrain and the medium. This equation is derived from the MacPherson-Srolovitz equation extended to anisotropic microstructures \cite{macpherson_von_2007,glicksman_mean_2009} with empirical assumptions about the relationship between the grain or subgrain size and its number of faces \cite{despres_mean-field_2020,zhang_three-dimensional_2018}. 
The translation of the capillary term to experimental microstructures is not straightforward since the mean boundary energy of the medium $\bar{\Gamma}$ can be defined at the local scale or for the whole microstructure. It was noticed, however, that the model predictions are not sensitive to this term regardless of its definition. Therefore, we make the simplifying assumption that $\Gamma_{(i)}=\bar{\Gamma}$, i.e that $a_{(i)}=1$. This is true on average since most subgrains grow within their deformed grains, where the orientation spread and thus the boundary network are homogeneous. In this case, \autoref{eq:growth-rate3} becomes:


\begin{equation}
\frac{dD_{\left(i\right)}}{dt}=4M_{\left(i\right)}\Gamma_{\left(i\right)}\left(\frac{8}{9\bar{D}}-\frac{1}{D_{(i)}}\right)
\label{eq:growth-rate3-h}
\end{equation}

This equation is almost identical to that derived by Hillert for the growth of grains in 3D \cite{hillert_theory_1965}. In this formulation, the growth rate of a grain or subgrain depends solely on its intrinsic properties and on the mean subgrain size, while the properties of the surrounding boundary network affect all grains and subgrains similarly (i.e. $a_{(i)}=1$). As discussed above, this assumption is reasonable for the present case, but its validity may need to be re-evalutated for more heterogeneous microstructures.

To update the grain and subgrain diameters, \autoref{eq:growth-rate3-h} is integrated using Euler’s method $\left[D_{(i)}\right]_{t+dt} = \left[D_{(i)}\right]_{t}+ \left[\frac{dD_{(i)}}{dt}\right]_t dt$. The mean grain and subgrain diameter $\bar{D}$ is also updated by performing a simple arithmetic mean of the list of diameters over the entire map. The model predictions are insensitive to the choice of $dt$ so long as the average increase in grain and subgrain volume per time increment remains below ~1\%. After each time increment, the smallest subgrains and those of negative radius are removed in order to maintain a constant total simulation volume \cite{despres_mean-field_2020}. This removal procedure affects all subgrains regardless of their parent deformed grains. Orientations whose grains and subgrains have fast growth rates will thus increase their fraction, while those associated with slower growth rates can subsist for some time but are ultimately removed from the simulation. 

As the microstructure evolves, recrystallized grains are identified, as in experiments, based on a threshold diameter $D_{( i)}\geq 30\mu m$. At high recrystallized fractions, the model predictions are almost insensitive to the definition of the threshold recrystallized grain diameter as kinetics and the texture are dominated by grains whose size has become several times that of the threshold.

\subsection{Determination of the input parameters}

To compute the growth rates given by \autoref{eq:growth-rate3-h}, the list of individual grain and subgrain diameters $D_{(i)}$, the mean diameter $\bar{D}$ and the mean boundary energies $\Gamma_{(i)}$ and mobilities $M_{(i)}$ must be known. The individual grain and subgrain diameters are obtained from the equivalent-area diameters measured on the experimentally measured orientation maps. It can be remarked that the grains and subgrains are assumed to grow in three dimensions and to possess volumes, even though their initial size is measured on two dimmensional sections. This considered, in first approximation, to have a negligible effect on the predictions since the apparent diameter of spherical grains and subgrains measured from two dimensional sections does not vary much on average from their real diameter in three dimensions \cite{mendelson_average_1969,gerlt_transfer_2018}. In addition, the cellular growth equations in 2D and 3D are almost identical in terms of their sensitivity to diameters and boundary properties \cite{despres_mean-field_2020}.


Next, the mean boundary properties are estimated from the boundaries separating each subgrain and its environment in the orientation maps. First, the boundary energy and mobility are assumed to be functions of the boundary disorientation angle $\theta$. The boundary energy $\gamma(\theta)$ is taken to obey the Read-Shockley equation \cite{read_dislocation_1950}:

\begin{equation}
\gamma\left(\theta\right)=
     \begin{cases} 
    \gamma_{c}\frac{\theta}{\theta_{c}}\left(1- \ln\frac{\theta}{\theta_{c}}  \right) & \text{if } \theta \leq \theta_c \\
    \gamma_c & \text{if } \theta > \theta_c
  \end{cases}
\label{eq:ReadSchockley}
\end{equation}

Where $\gamma_c$ is a constant and $\theta_c$ is a cut-off angle set to 15$^\circ$ to simulate a high angle boundary. The boundary mobility $\mu\left(\theta\right)$ is set to follow the empirical relation \cite{humphreys_unified_1997,huang_subgrain_2000}:

\begin{equation}
\mu \left(\theta\right)=\mu_{c}\left(1- e^{-B\left(\frac{\theta}{\theta_{c}}\right)^{\eta}}  \right)
\label{eq:Huang}
\end{equation}

Where $\mu_c$ is a constant, $B=5$ and $\eta=4$ \cite{humphreys_unified_1997,huang_subgrain_2000}. The constants $\gamma_c$ and $\mu_c$ have no influence on the prediction of the recrystallization texture, but the kinetics scale with their magnitude. The choice of these equations leads, in the simulation, to the development of recrystallized grains with boundaries of high disorientation angles \cite{humphreys_unified_1997-1,despres_mean-field_2020}. This outcome is consistent with the fact that recrystallized grains are mostly separated from the surrounding deformed grains and from other recrystallized grains by high-angle boundaries with no particular orientation relationship. 

\begin{figure}[htbp]
    \centering
\begin{minipage}[t]{0.49\textwidth}
\centering
   \includegraphics[width=\textwidth]{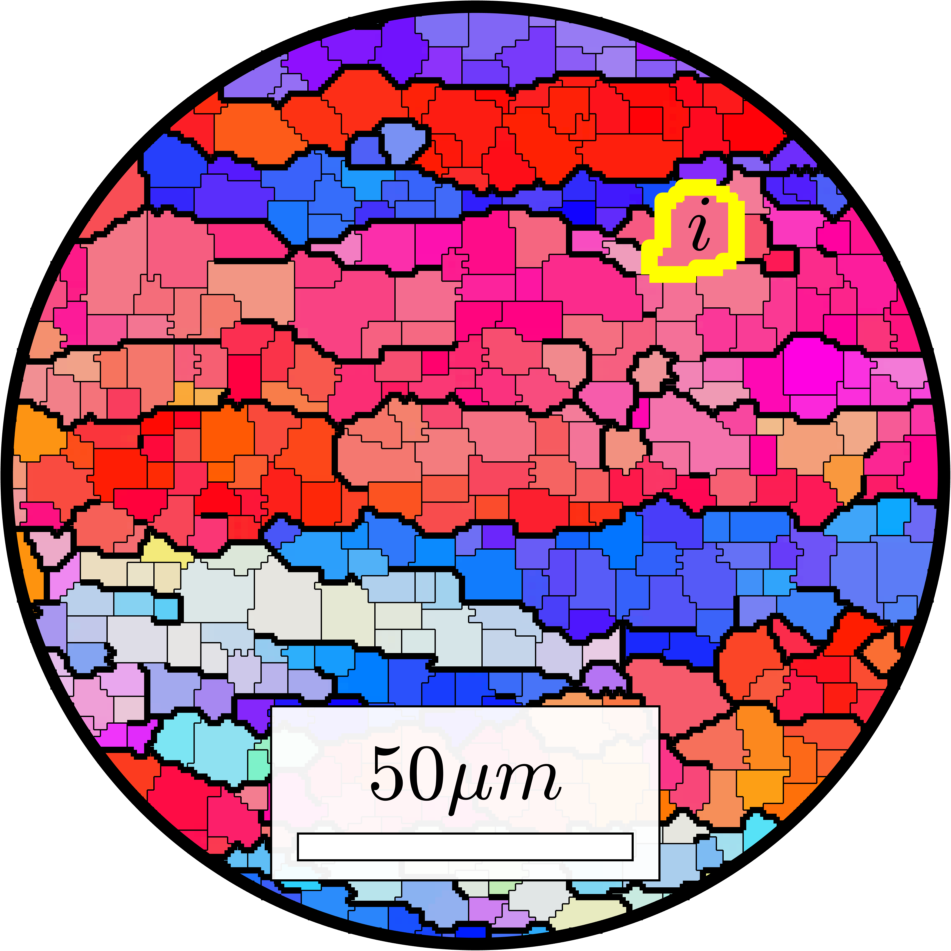}
\end{minipage}
\begin{minipage}[t]{0.49\textwidth}
\centering
   \includegraphics[width=0.35\textwidth]{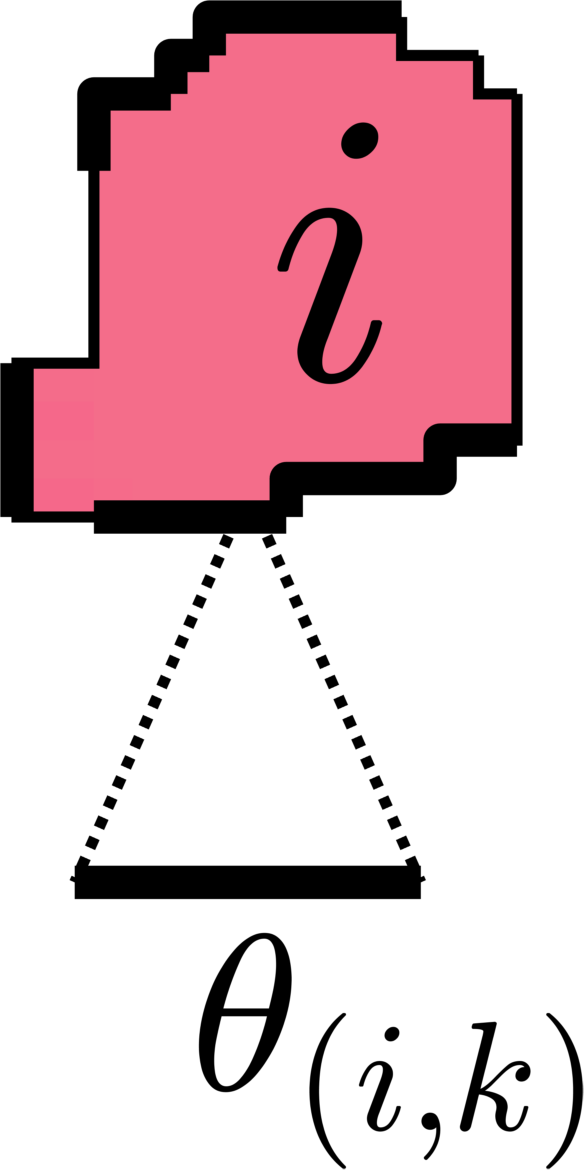}
   
\end{minipage}
\caption{Illustration of the measurement of disorientation angles at the boundaries of a given subgrain $i$. The unit segment of disorientation angle $\theta_{(i,k)}$ has the length of the EBSD step size.}
\label{fig:ebsd-example}
\end{figure}

Having set the boundary energy and mobility laws, one can then calculate the mean boundary properties of subgrains. The method is illustrated in \autoref{fig:ebsd-example}. The mean boundary energy of a subgrain $i$ is given by the aritmetic mean:

\begin{equation}
	\Gamma_{\left(i\right)}=\sum_{k=1}^{n_{(i)}}\gamma\left(\theta_{(i,k)}\right)/n_{(i)}
	\label{eq:energy}
\end{equation}

Where $n_{(i)}$ is the number of unit segments defining the subgrain boundary and $\theta_{(i,k)}$ is the disorientation angle of the $k$-th segment.

Similarly, the mean boundary mobility assigned to the subgrain is given by:

\begin{equation}
	M_{\left(i\right)}=\sum_{k=1}^{n_{(i)}}\mu\left(\theta_{(i,k)}\right)/n_{(i)}
	\label{eq:mobility}
\end{equation}

In our previous work \cite{despres_mean-field_2020}, the mean boundary energy and mobility were calculated by second order Taylor series expansions of the boundary energy and mobility laws around the means and variances of the boundary disorientation angle distributions. Except small discrepancies linked to sampling, the Taylor series leads to identical results as the arithmetic mean if the means and variances used for the Taylor series are calculated at the scale of each individual subgrain. The arithmetic mean is prefered here as, from a theoretical point of view, Taylor series expansions must converge towards the arithmetic mean when the order tends to $+\infty$.

Finally, it is convenient to also calculate the mean boundary disorientation of individual grains and subgrains even though this is not an input parameter of the model. This parameter has some utility for discussing the model prediction as both boundary energy and mobility depend on disorientation angle. It is given by:

 \begin{equation}
	\Theta_{\left(i\right)}=\sum_{k=1}^{n_{(i)}}\theta_{(i,k)}/n_{(i)}
\end{equation}

Note that, for simplicity, the boundary properties are not updated with time. This assumption is expected to have a negligible effect on the texture prediction since it was noticed in the orientation maps that even though the mean boundary properties of grains and subgrains evolve during recrystallization, orientations associated to grains and subgrains of high boundary energy and mobility maintain these high values throughout recrystallization. In addition, the differences in boundary properties between grains and subgrains located within deformed grains and those adjacent to deformed grain boundaries are expected to remain similar throughout recrystallization. On the one hand, those within deformed grains will mainly retain boundaries of low disorientation angles since they grow in regions with weak orientation gradients. On the other hand the high disorientation angles of grains and subgrains at deformed grain boundaries are geometrically necessary to rotate from orientations on the one side to those on the other side of the boundaries. 

\subsection{Algorithm}

The input parameters to the model are obtained from the experimentally measured orientation maps of the deformed microstructure. The initial simulation volume is obtained by summing the volumes of all subgrains in the input microstructure. The subgrains volumes are calculated from their diameters assuming spherical particles. Then a time iteration loop is started, with the following sequence executed between times $t$ and $t+dt$:

\begin{enumerate}
\item Identify the recrystallized grains, i.e. the subgrains with $D_{(i)}\geq 30\mu m$.
\item Calculate the growth rate of each grain and subgrain using \autoref{eq:growth-rate3-h}.
\item Integrate the growth rates over a time increment to update the grain and subgrain diameters. The new diameters are representative of the microstructure at time $t + dt$.
\item Remove the subgrains of negative diameter and the smallest subgrains of positive diameter so as to minimize the difference between the initial microstructure volume and the sum of grain and subgrain volumes at $t + dt$.
\item Update the average diameter $\bar{D}$.
\end{enumerate}

At each time step, the recrystallized fraction and the orientation of the recrystallized grains are known. Therefore, both the recrystallization kinetics and texture can be predicted.




\section{Results and discussion}

\subsection{Kinetics and texture prediction}

\begin{figure}[htbp]
\centering
\includegraphics[width=0.45\textwidth]{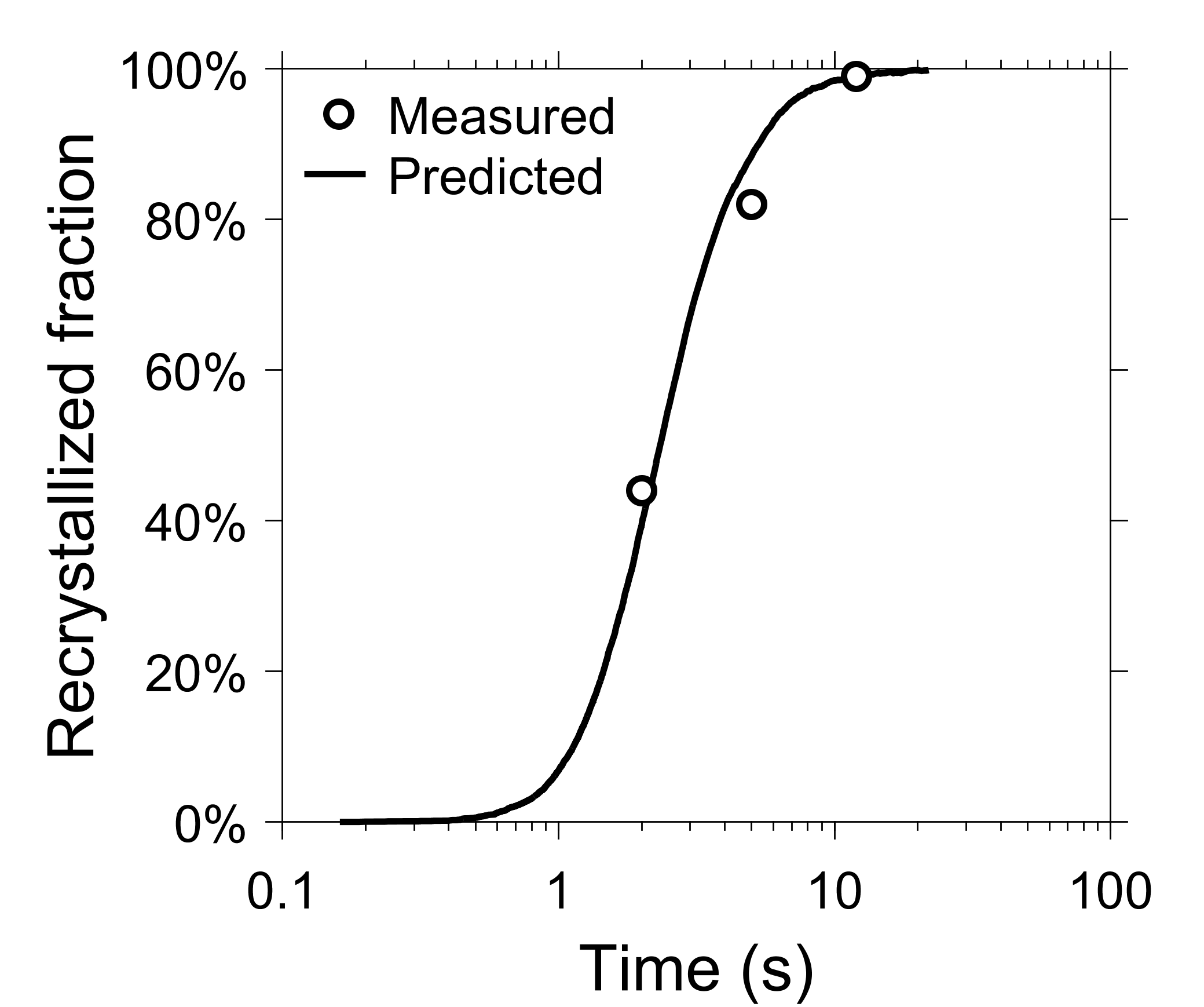}
\caption{Experimental and simulated recrystallization kinetics. The energy and mobility factors have been set to $\gamma_c=0.8\times10^{-12}J.\mu m^{-2}$ and $\mu_c=1\times10^{14}\mu m^4.J^{-1}.K^{-1}$.}
\label{fig:kinetics}
\end{figure}

\autoref{fig:kinetics} compares the recrystallization kinetics predicted by the model to the experimental datapoints. The predicted kinetics follows the sigmoidal shape characteristic of recrystallization kinetics, and fits the datapoints when setting $\gamma_c=0.8J.m^{-2}$ and $\mu_c=1\times10^{-10}m^4.J^{-1}.K^{-1}$. The time required to reach the annealing temperature has not been considered in the simulation. In previous models for recrystallization of ferritic steels, $\gamma_c$ usually ranges from 0.65 to 0.8$J.m^{-2}$ \cite{jacquet_etude_2013,sinclair_effect_2007}. For $\mu_c$, the Arrhenius law suggested by Jacquet \cite{jacquet_etude_2013} gives $\mu_c=0.96\times 10^{-10}m^4.J.s^{-1}$ at 1100$^\circ$C\footnote{To simulate the recrystallisation kinetics in a hot rolled AISI445 ferritic stainless steel using a Bailey-Hirsch type of model, Jacquet suggested $\mu_c=\mu_{c,0}exp(-Q/RT)$, with $\mu_{c,0}=5\times 10^{-6}m^4.J.s^{-1}$, $Q=124kJ.mol^{-1}$ and $R$ the ideal gas constant \cite{jacquet_etude_2013}. These parameters were obtained from grain growth experiments.}. The order of magnitude of the fitting constants is thus in good agreement with that found in the literature. The ability of this model to predict the recrystallization kinetics had been suggested previously on the basis of full-field and mean-field simulations \cite{despres_mean-field_2020}. The present result can be considered as a first successful case using experimental measurements as input.


\begin{figure}[htbp]
\centering
\subfigure[]{\includegraphics[width=0.32\textwidth]{odf-0s_weighted.png}}
\subfigure[]{\includegraphics[width=0.32\textwidth]{odf-12s_weighted.png}}
\subfigure[]{\includegraphics[width=0.32\textwidth]{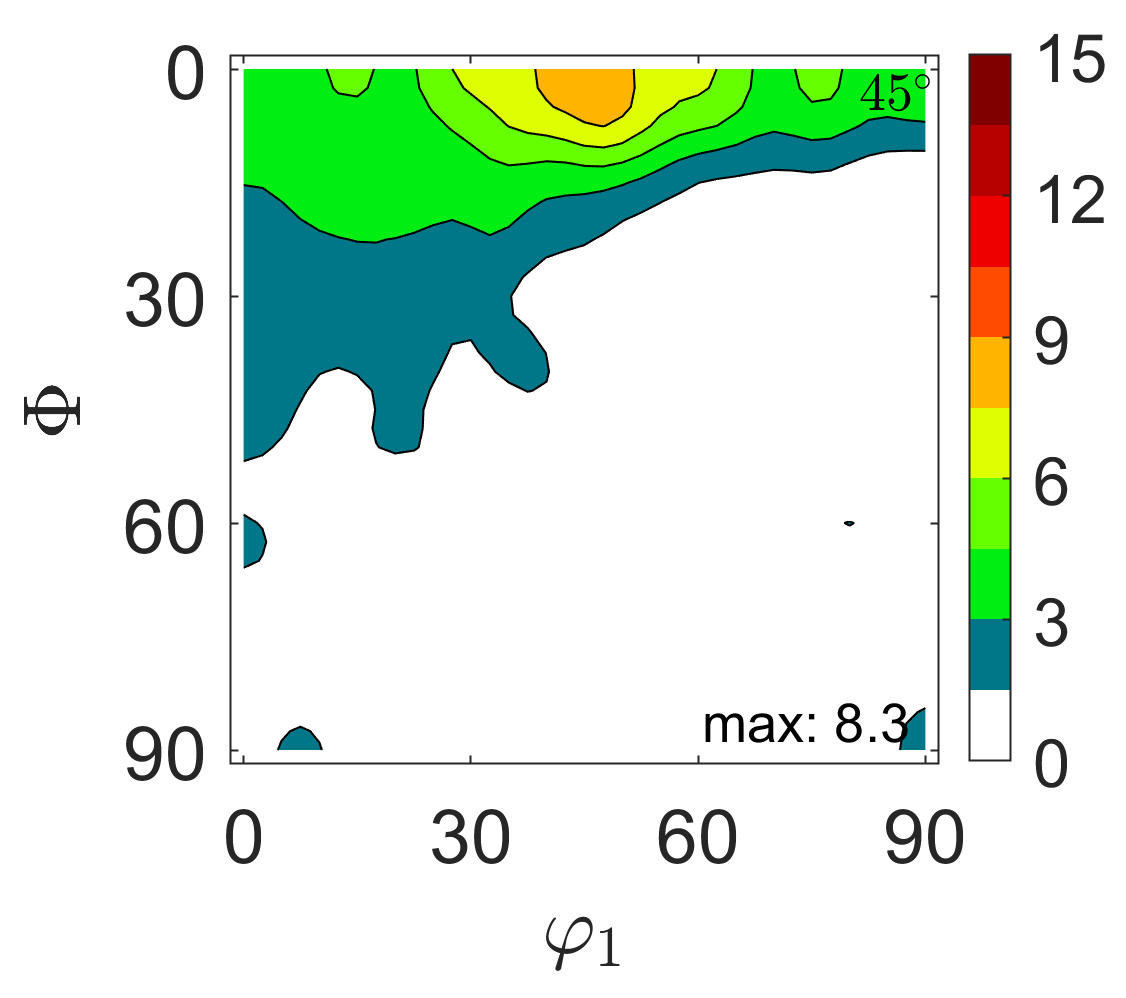}}



\caption{ODFs sections in the $\varphi_2=45^\circ$ section of the Euler space for (a) the experimentally measured deformation texture, (b) the experimentally measured texture at 97\% of recrystallized fraction (after 12s annealing), (c) the predicted texture at 97\% of recrystallized fraction. The scalebar is in multiple of random (m.r.d).}
\label{fig:ODF-comparison}
\end{figure}

\autoref{fig:ODF-comparison} compares the experimental and predicted recrystallization textures at 96\% recrystallized fraction. The measured deformation texture is shown again in \autoref{fig:ODF-comparison}a to illustrate the differences with the measured and predicted recrystallization textures in \autoref{fig:ODF-comparison}b and c. The prediction captures the weakening of the texture strength with recrystallization as well as the development of the cube orientation $\{001\}\langle100\rangle$ as the strongest texture component. With this way of representing the texture, the most visible discrepancy between the experimental and predicted texture is the absence of a local maximum near the rotated cube orientation $\{110\}\langle100\rangle$. Overall, however, the evolution of this component is well captured if one compares with its strength in the deformed state.


 
Recall that the recrystallization texture was simulated starting from a list 1.8$\times$10$^5$ subgrains measured in the deformed microstructure. Out of these, about 750 reached the threshold recrystallized grain size at the end of recrystallization \footnote{If the simulations are performed assuming 2D growth (using equations developed in ref. \cite{despres_mean-field_2020}), the number of recrystallized grains reaches several thousands for an almost identical texture prediction.}. Including a larger number of initial subgrains did not lead to significant differences in the predicted texture. It can be remarked that this number of measured subgrains in the deformed microstructure can be acquired rapidly with modern EBSD systems. Thus sampling of the inital deformed microstructure is not a major constraint for the simulation. 







\subsection{Dynamics of the recrystallization texture development}

\begin{figure}[htbp]
\centering
\includegraphics[width=0.5\textwidth]{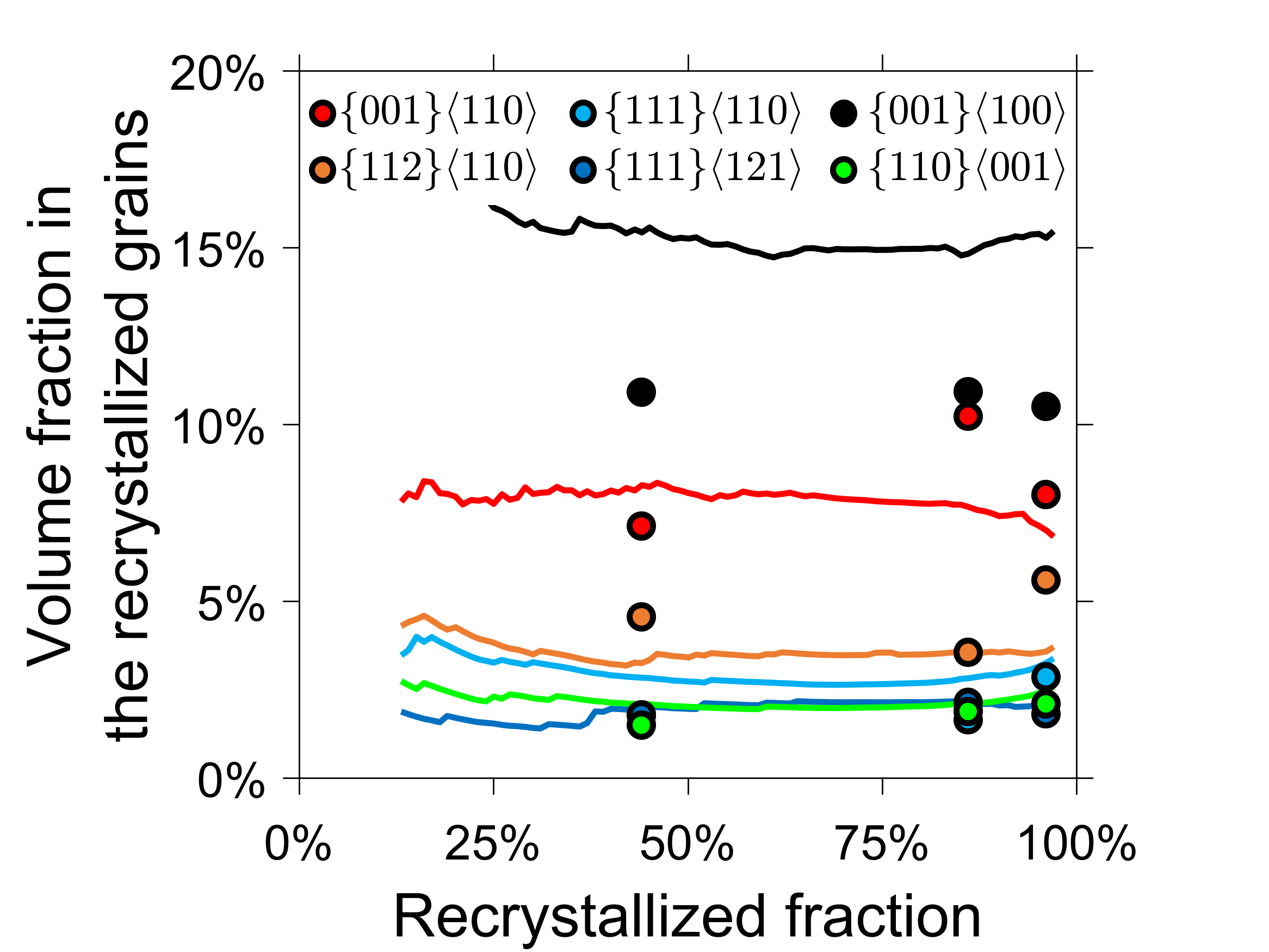}
\caption{Volume fraction of recrystallized grains within 15$^\circ$ from some low-index texture components as a function of the recrystallized fraction. The experimental measurements are shown in dots and the predictions in lines.}
\label{fig:fractions}
\end{figure}

\autoref{fig:fractions} shows the 
volume fraction of several low-index texture components in the recrystallized grains as a function of the recrystallized fraction. The measurements corresponding to the 2s, 5s and 12s annealing conditions are shown in dots while the model predictions are shown as lines. The volume fraction is taken here as another way of quantitatively comparing the textures. Plotted in this way, the volume fraction of the cube orientation $\{001\}\langle100\rangle$ seems overpredicted. However, the discrepancy is not more than 50\% for all conditions, which is satisfactory given the complexity of the mechanisms simulated and the number of simplifying assumptions. Besides, the general evolution of this component is well captured by the model as its volume fraction increases significantly between the deformed and the recrystallized state, as was found in the experiment (see \autoref{tab:volume-fractions}).

Both the experimental points and the model predictions show that the volume fractions of recrystallization texture components are set at low recrystallized fractions and do not evolve much with the progress of recrystallization. 
It would be tempting to relate this behaviour to a phenomenon of `oriented nucleation', where the recrystallization texture is dominated by the differences in nucleation rate of recrystallized grains as a function of their orientation \cite{humphreys_recrystallization_2017,doherty_current_1997}.  However, we prefer not to use this term as the distinction between nucleation and growth is not well suited to the present case. Indeed, according to the model, the texture change results from the competitive growth of subgrains, and nucleation of recrystallized grains is controlled by a size threshold that has no fundamental meaning.

The stability of the recrystallization texture throughout recrystallization is not necessarily surprising if one considers the topological aspects of recrystallization in this material and the theoretical conditions giving rise to recrystallized grains. According to \autoref{eq:growth-rate3-h} and in agreement with the classic point of view \cite{rollett_growth_1997,humphreys_unified_1997-1}, the subgrains with boundaries of highest energy and mobility have the fastest growth rates and, as a consequence, are the most likely to reach the threshold recrystallized grain size. As boundary energy and mobility are related to the boundary disorientation angle, subgrains surrounded by high-angle boundaries have the highest chances of forming recrystallized grains. As recrystallized grains grow towards their neighbouring deformed grains (see \autoref{fig:ebsd-maps}), their boundaries mainly retain high disorientation angles. For example, if one considers that the orientation of the recrystallized grains and of the deformed microstructure are spatially uncorrelated, a recrystallized grain of exact rotated cube orientation has a 96\% chance of sharing a boundary of disorientation angle above 10$^\circ$ with the deformed texture. This probability is even higher for the other texture components since they are present in lower fraction in the deformed texture. The stability of the recrystallized grain boundary properties throughout recrystallization helps to explain why the relative amount of each recrystallization texture component does not change after the appearance of the first recrystallized grains. As the early stages of the microstructure evolution determine most of the texture change, an examination of the initial distribution of subgrain properties should suffice to explain the origin of the recrystallization texture in this material.

\subsection{Origin of the recrystallization texture}



\begin{figure}[htbp]
\centering
\subfigure[]{\includegraphics[width=0.45\textwidth]{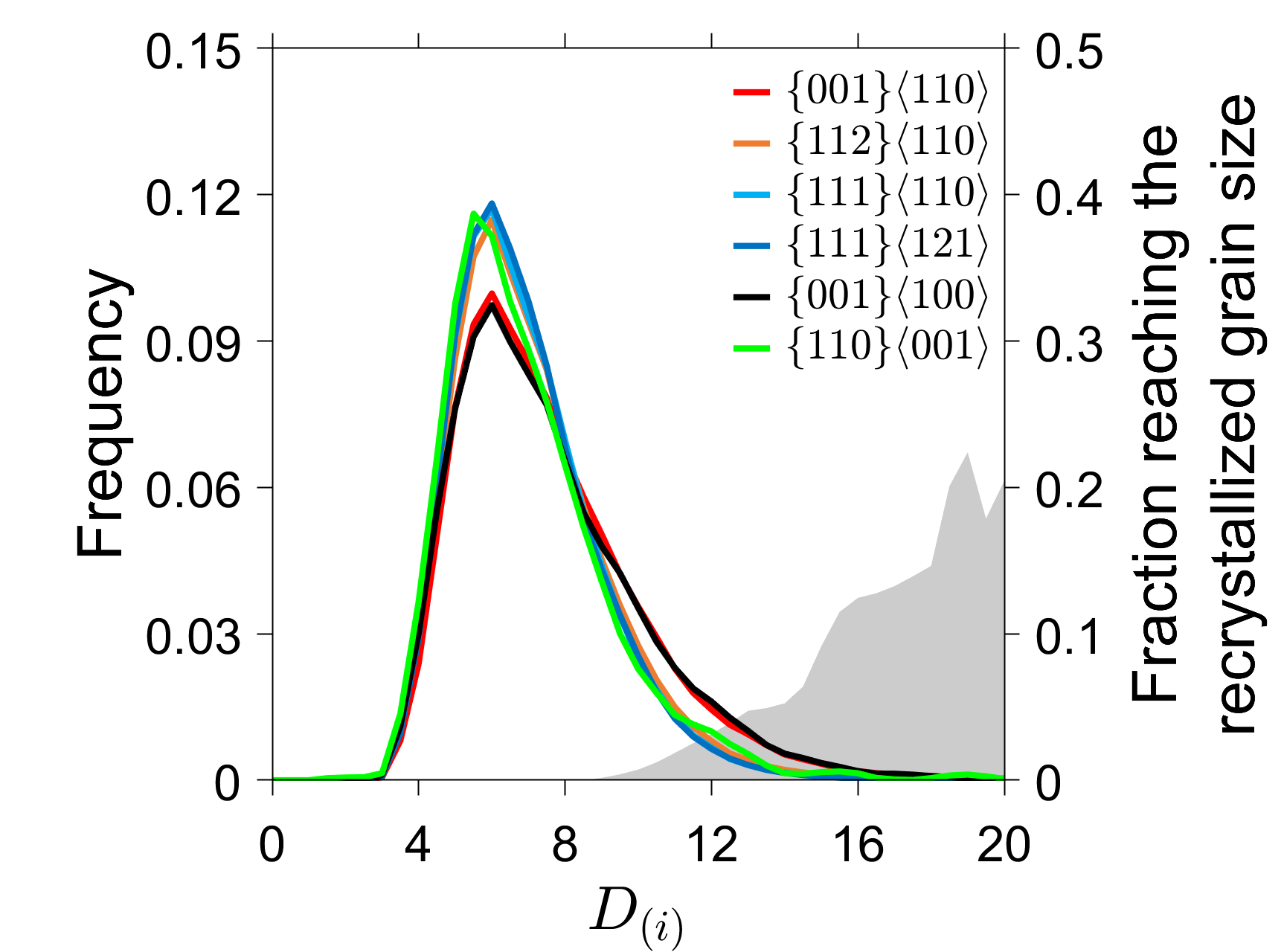}}
{\includegraphics[width=0.45\textwidth]{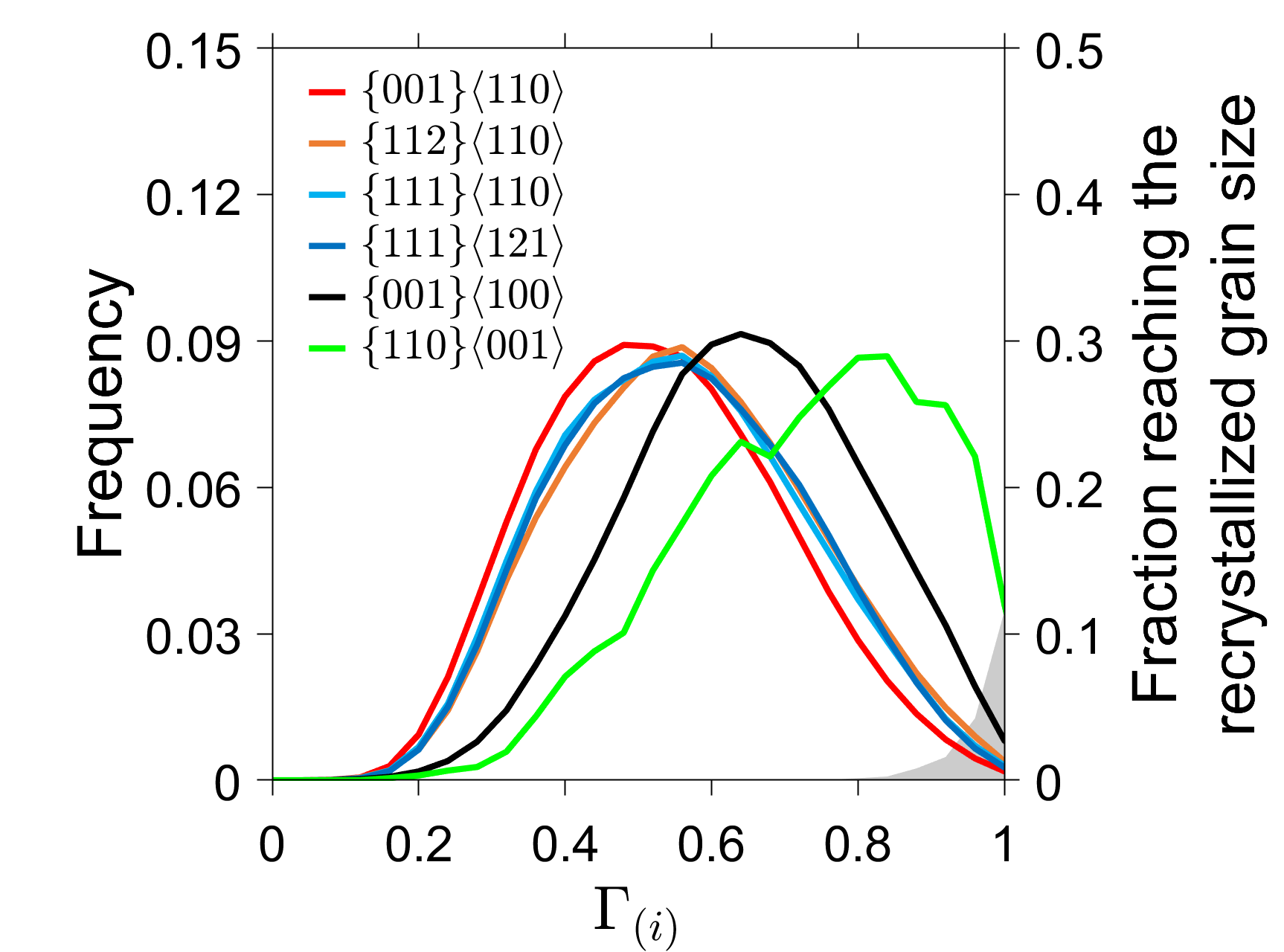}}

{\includegraphics[width=0.45\textwidth]{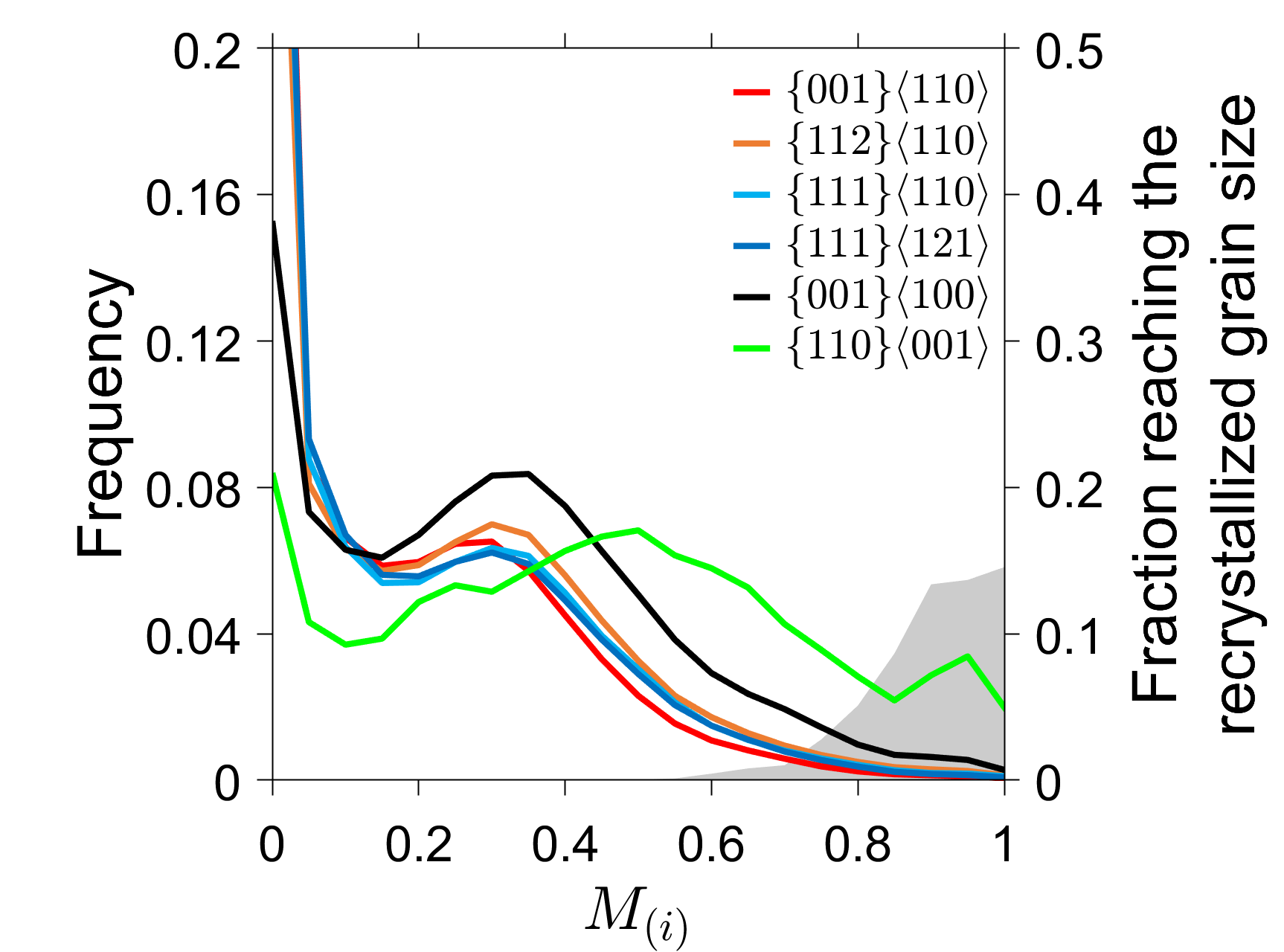}}
{\includegraphics[width=0.45\textwidth]{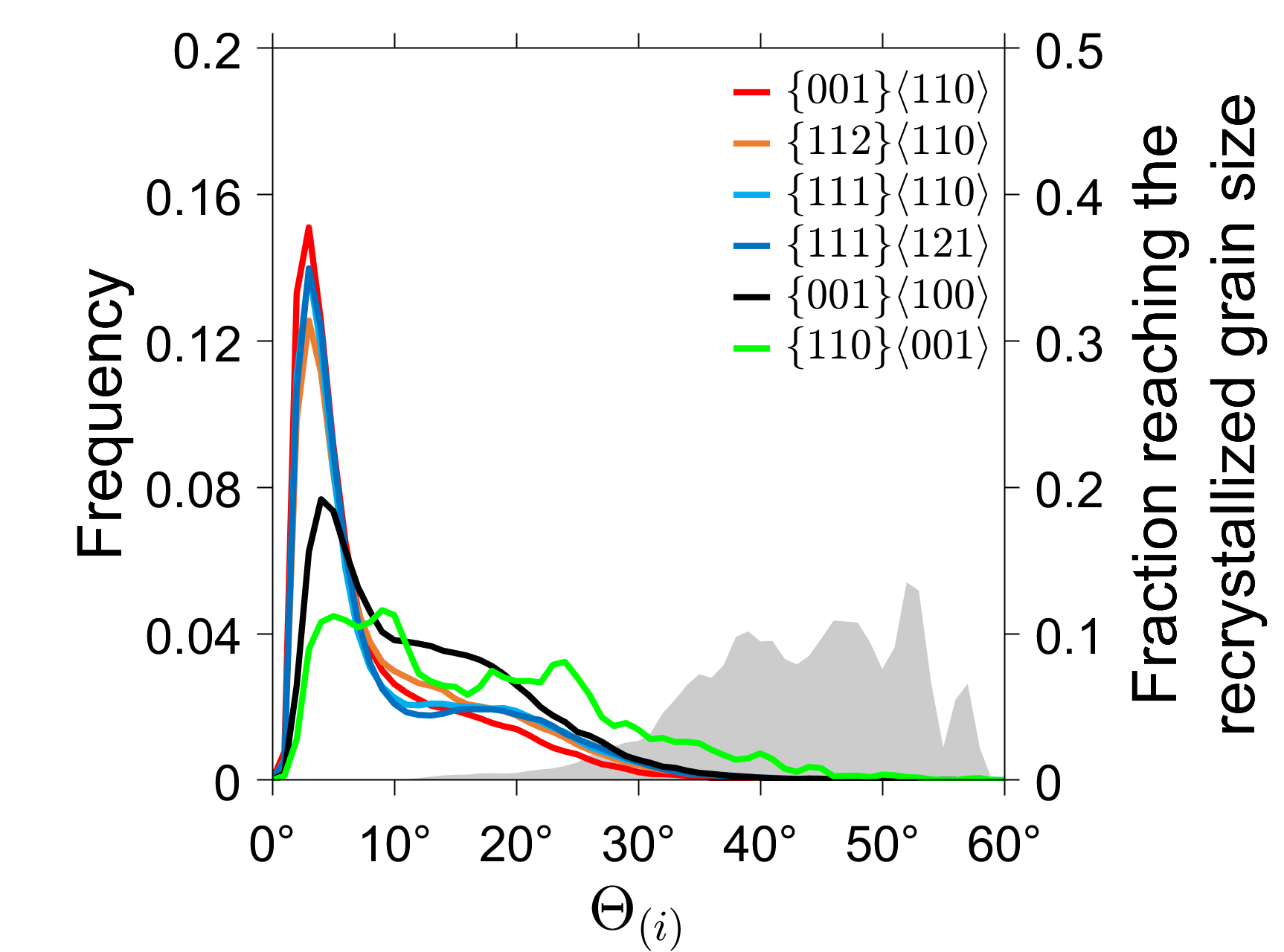}}

\caption{Distribution of subgrain properties in the deformed microstructure as a function of their orientation (in coloured lines). The grey areas represent the fraction of the initial subgrains having reached the threshold recrystallized grain size at a recrystallized fraction of 97\%. a) subgrain diameter, b) mean boundary mobility, c) mean boundary energy, d) mean boundary disorientation angle. }
\label{fig:subgrain-prop}
\end{figure}

In the model, the distribution of subgrain properties and the initial number of subgrains belonging to each texture component determine the recrystallization texture development. To simplify the analysis, we evaluate separately the effect of the different subgrain parameters on the recrystallization texture development. For this, \autoref{fig:subgrain-prop} shows the distributions of subgrain diameters, mean boundary energy, mean boundary mobility and mean boundary disorientation angle for several low-index components measured from the orientation maps of the deformed microstructure. On the same plots, the shaded areas represent the fraction of subgrains reaching the recrystallized grain size at 97\% recrystallized fraction. \autoref{fig:subgrain-prop}a to c confirm that the probability for a subgrain to reach the recrystallized grain size increases with the subgrain diameter, mean boundary energy and mobility. As boundary energy and mobility are related to the boundary disorientation, \autoref{fig:subgrain-prop}d shows that the probability of turning into a recrystallized grain also increases as a function of the mean disorientation angle, in agreement with the common perception \cite{rollett_growth_1997,humphreys_unified_1997-1}. Note that even if subgrains of extreme properties have the most chance of reaching the recrystallized grain size, they do not necessarily represent the majority of the recrystallized grains since they make up a very small fraction of the deformed microstructure. 

Interestingly, subgrains of the $\alpha$ fibre, i.e. the $\{001\}\langle110\rangle$, $\{112\}\langle110\rangle$, $\{111\}\langle110\rangle$ orientations do not exhibit particularly extreme boundary properties (\autoref{fig:subgrain-prop}b-d), even though they form the one of main components of the recrystallization texture. Thus, their presence in the recrystallized state can be attributed to their high fraction in the deformed state. This interpretation is supported by the fact that  the volume fraction of these components decreases with recrystallization (see \autoref{tab:volume-fractions}). By contrast, subgrains of the cube orientation $\{001\}\langle100\rangle$ possess higher boundary energy and mobility (\autoref{fig:subgrain-prop}b and c), which gives them a higher probability of turning into recrystallized grains and explains why this component develops preferentially. The general weakening of texture with recrystallization can be explained using the same reasoning since components such as the Goss orientation $\{110\}\langle001\rangle$ possess boundary properties that allow them to develop to a much higher extent than the components that are much more present in the deformed texture.

\subsection{Origin of the anisotropy in subgrain properties}

The present analysis suggests that the anisotropy in subgrain properties and the deformation texture prior to recrystallization control the recrystallization texture development. While the principal aim of this work is not to explain how to control the anisotropy in subgrain properties, it is interesting to point out that this originates from the material's behaviour during prior deformation. 


To begin with, \autoref{fig:subgrain-prop}a shows a slight increase in subgrain size along the $\alpha$ fibre from the $\{111\}\langle110\rangle$ to the $\{001\}\langle110\rangle$ orientations. This evolution is well documented in the literature, and is attributed to differences in slip activity as a function of orientation \cite{hutchinson_deformation_1999,dillamore_oriented_1967,kestens_modelling_1997}.  The anisotropy in subgrain size is however less significant than that observed in cold rolled materials \cite{dillamore_oriented_1967}, because the high strain rate sensitivity of hot rolled materials homogenizes the plastic response of grains as a function of orientations \cite{hutchinson_deformation_1999}. In \autoref{fig:subgrain-prop}b-d, the boundary properties of the main deformation texture components (i.e. the $\{001\}\langle110\rangle$, $\{112\}\langle110\rangle$, $\{111\}\langle110\rangle$ and $\{111\}\langle121\rangle$ orientations) appear quasi self-similar, most likely for the same reason

By contrast, the exceptionally high boundary properties of the cube orientation $\{001\}\langle100\rangle$ and Goss orientation $\{110\}\langle001\rangle$ can be explained by the fact that these are metastable orientations under plane strain compression of bcc materials and develop in low fraction in the deformed microstructure \cite{dillamore_transition_1972}. Such structures have been called transition bands by Dillamore \emph{et al.} \cite{dillamore_transition_1972}, although the term is rarely used for ferritic steels. From their point of view, the presence of locally strong orientation gradients and the low fraction of these components should ensure that they possess with the rest of the microstructure more high-angle boundaries than the other texture components. This assertion is supported by \autoref{fig:subgrain-prop}d, where the cube $\{001\}\langle100\rangle$ and Goss $\{110\}\langle001\rangle$ orientations are shown to possess a higher fraction of boundaries with angles above approximately 10$^\circ$ than the other investigated components.


At constant rolling temperature, the anisotropy in subgrain properties is unlikely to change radically since it is imposed by the strain rate sensitivity, which is temperature dependent \cite{jacquet_etude_2013}. This seems to be confirmed by measurements of subgrain properties in the same material deformed at 1100$^\circ$C with a 50\% reduction rate \cite{despres_origin_2019}. Thus, to radically change the recrystallization texture, one would have to play on the rolling temperature. Warm and cold deformation are known to induce more strain heterogeneities and a larger fraction of high-angle boundaries in $\gamma$ fibre grains \cite{barnett_role_1998,despres_contribution_2020}. This is expected to help lowering the fraction of $\alpha$ fibre orientations in the as-recrystallized state \cite{zhang_effects_2011,barnett_role_1998}. 

\subsection{Merits and potential of the model}

The principal merit of this model is its simple expression of the relationship between subgrain parameters in the deformed state and the texture development during recrystallization. The parameters which drive recrystallization, namely the distributions of subgrain size and boundary property, can be measured directly on orientation maps obtained from experiments. Therefore, one can use the model to investigate the characteristics of the subgrains in the deformed microstructure which give rise to recrystallized grains. This can be done in a statistical way, as presented in \autoref{fig:subgrain-prop}. But one may also use the model results to identify the most favourable sites for the formation of recrystallized grains on the orientation maps of the deformed microstructure. For example, in the material investigated here, recrystallized grains are expected to form near the pre-existing deformed grain boundaries since this is where most high-angle boundaries are located. In a material exhibiting a higher degree of intragranular heterogeneities, one may expect the model to locate more frequently the origin of recrystallized grains in the interior of the deformed grains. This analysis is not possible with other models of the literature where the predicted texture is tuned by parametrically weighting the contribution of different regions of the microstructure (e.g. transition bands, deformed grain boundaries) to the formation of recrystallized grains  \cite{sebald_modeling_2002,sidor_modeling_2011,zecevic_modelling_2019}.


One can also view this model as part of an effort to develop physically sound approaches for predicting the texture evolution of wrought materials during thermomechanical processing. For example, its input parameters could be measured from synthetic orientation maps obtained by full-field crystal-plasticity simulations instead of experimentally measured maps. Another possibility would be to use the outcomes of mean-field crystal plasticity simulations to estimate the subgrains properties. As discussed previously \cite{despres_mean-field_2020}, this is made possible by the fact that the outputs of mean-field crystal-plasticity models are very much similar to the input parameters of the recrystallization model. For example, distributions of disorientation angles can be generated from the intragranular orientation spreads calculated by the model of Zecevic \emph{et al.} \cite{zecevic_modeling_2018,zecevic_modelling_2019}. Subgrain sizes are not usually an output of crystal plasticity models, but these can be estimated, in first approximation, from the variations of Taylor factors \cite{hutchinson_deformation_1999}. Using mean-field models to feed recrystallization models seems a particularly interesting strategy to perform through-process modelling of texture evolution as it would retain a high computational efficiency.

As a final remark, it is worth pointing out that the model assumptions have been selected after a careful examination of the microstructures shown in \autoref{fig:ebsd-maps}. As a result, the model is expected to be most suited to the recrystallization of high-stacking fault energy materials, where the deformed microstructure can reasonably be considered as a subgrain network with low dislocation density. The proposed implementation is however flexible since its formalism of cellular growth is comparable to the models existing in the literature \cite{hurley_modelling_2003,zurob_quantitative_2006} . Mathematical expressions accounting for the effect of other microstructural features such as orientation gradients \cite{pantleon_retrieving_2005,lefevre-schlick_activation_2009}, stored energy due to tangled dislocations \cite{zurob_quantitative_2006}, and precipitate pinning \cite{humphreys_unified_1997} have been proposed elsewhere. Including these effects could be readily accomplished by altering some of the assumptions presented here.

\section{Conclusion}

A model was developed to predict the static recrystallization texture development during annealing of deformed polycrystals. The model assumes that recrystallization arises from the competitive growth of subgrains, driven by the anisotropy of subgrain properties. One of its distinctive features compared to other models of the literature is its ability to use as input parameters experimental measurements of the subgrain properties. This approach allows one to establish a direct relation between the state of the deformed microstructure and the recrystallization texture. The recrystallization kinetics can also be predicted.

The model predictions are in good agreement with the experimental measurements of a hot rolled and recrystallized ferritic stainless steel sheets. The results indicate that the strength of the texture and its main components develop at the early stages of recrystallization and are mostly determined by the initial distributions of subgrain properties. Due to their large initial fraction, the main components of the deformation texture compose the dominant fraction of the recrystallization texture even though they are not particularly in favourable growth conditions. The weakening of the texture with recrystallization is, on the other hand, explained by the rapid growth of the minor components. To minimize the fraction of deleterious $\alpha$ and cube fibre orientations in the recrystallized grains, one must minimize their presence in the states before recrystallization. Another possibility is to decrease the rolling temperature to induce more anisotropy in the subgrain properties.

The model assumptions have been chosen after a careful examination of the microstructural features of the material in its deformed and recrystallized states. As a result, the implementation proposed here is most suited to the simulation of recrystallization in high-stacking fault energy materials, whose deformation microstructure can be considered as a population of subgrains. It is also expected to perform well in cases where the recrystallization texture is determined by the early stages of the microstructure evolution, since the boundary properties are set constant with time. The model is however quite flexible, and its assumptions could be relaxed or modified to account for other microstructural features such as orientation gradients, stored energy due to tangled dislocation or precipitate pinning. 


%


\section*{Supplementary material}

The model input file is provided in supplementary material.

\section*{Ackowledgements}

This work was financially supported by APERAM. The authors express their thanks to Francis Chassagne from APERAM for his contribution to this project before his retirement.


\bibliographystyle{elsarticle-num}
\bibliography{Biblio}

\appendix
\begin{appendices}

\section{Implementation of the Kuwahara filter}
\label{app:kuwa}

The Kuwhara filter is an edge-preserving smoothing filter, whose application to orientation maps has first been developed by Humphreys \emph{et al.} \cite{brough_optimising_2006,humphreys_orientation_2001,humphreys_review_2001}. On raw orientation maps, the array of pixels surrounding each individual pixel is divided in four sub-arrays (north-east, north-west, south-west, south-east), and the individual pixel is reassigned the mean orientation of the sub-array having the least variance in orientation. According to Brough \emph{et al.} \cite{brough_optimising_2006}, using this filter allows the noise level to be reduced to below 0.3$^\circ$ while conserving the subgrain boundaries. As recommended \cite{humphreys_orientation_2001},
3 passes were performed with an array of 5x5 and subarrays of 3x3 pixels. Filtering was performed with the scripts implemented in MTEX \cite{noauthor_mtex_nodate}.

\begin{figure}[htbp]
\centering
\subfigure[]{\includegraphics[width=0.3\textwidth]{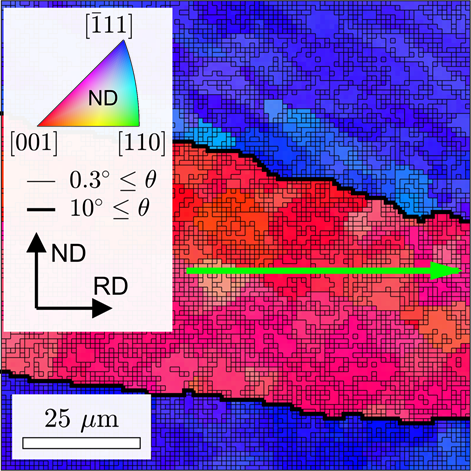}}
\subfigure[]
{\includegraphics[width=0.3\textwidth]{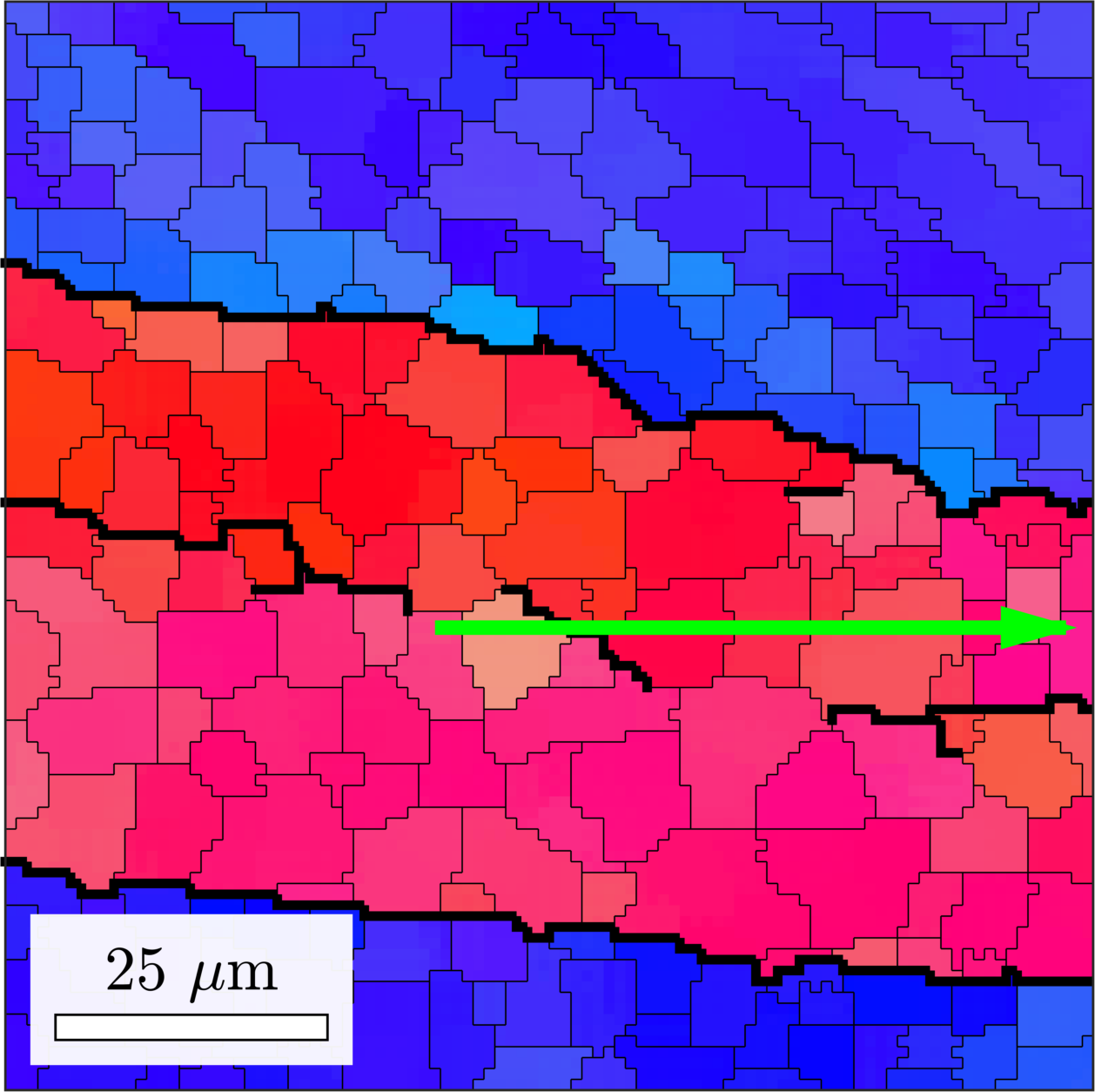}}
\subfigure[]
{\includegraphics[width=0.3\textwidth]{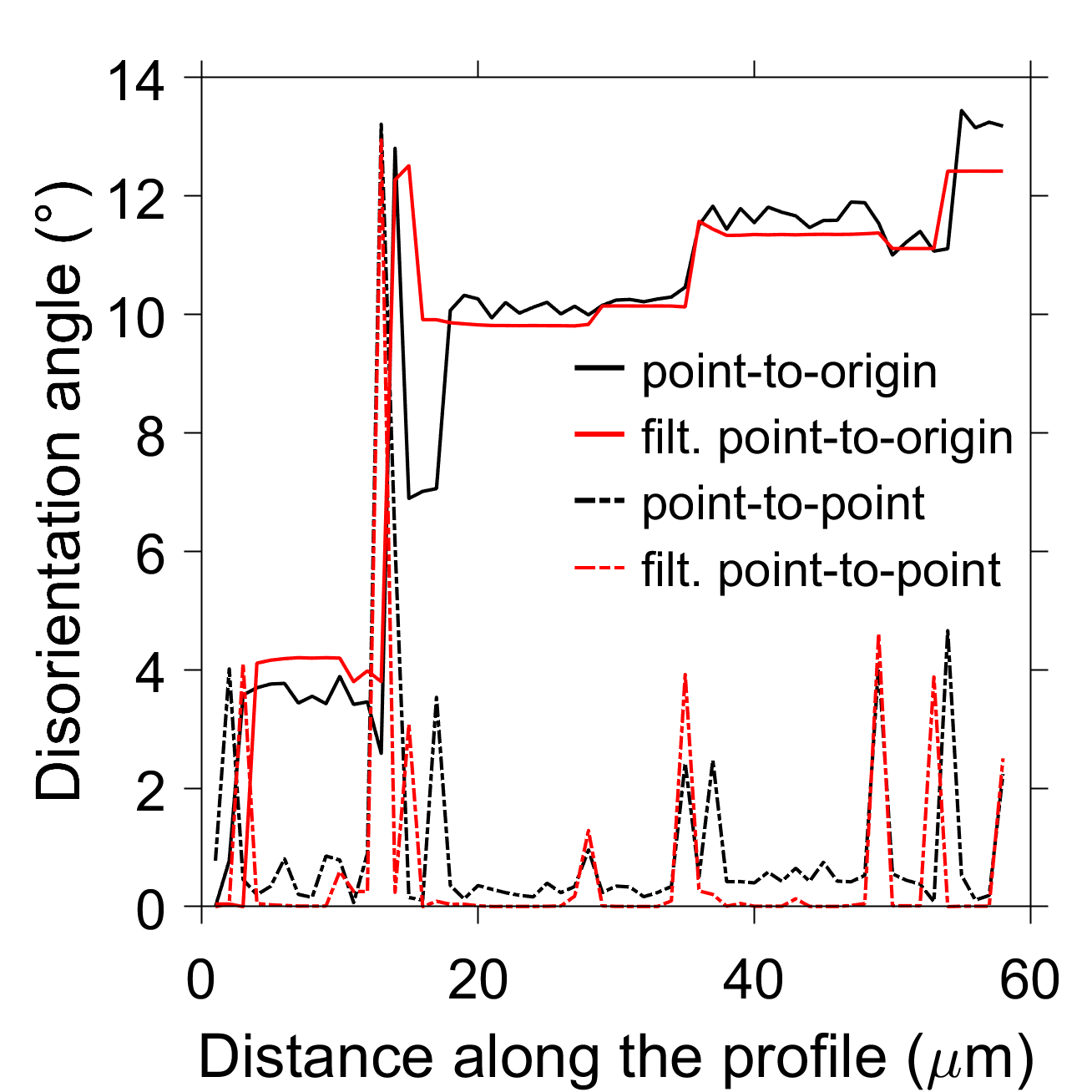}}

\caption{The effect of Kuwahara filtering on subgrain characterization in the deformed microstructure. (a) raw ND-IPF map, (b) filtered ND-IPF map, (c) disorientation profile along the green profile.}
\label{fig:illustration}

\end{figure}

 \autoref{fig:illustration}a and b illustrate the effect of filtering on a portion of IPF map of the deformed microstructure. \autoref{fig:illustration}c shows that filtering preserves the orientation gradient (point-to-origin disorientations) as well as the boundary disorientation angles which are already above a few degrees in the raw orientation map (note that these boundaries can shift horizontally and vertically of a few pixels). On the other hand, the minimum level of boundary disorientation angle falls to 0 in the filtered map. Modifying the threshold disorientation angle for subgrains is not expected to change significantly the main results of this study. For example, if a 1$^\circ$ angle instead of 0.3$^\circ$ is choosen to define subgrain boundaries, the average subgrain size increases of only 8.5\%, and the anisotropy in subgrain properties remains similar.

%
%
%
%
%


\end{appendices}
\end{document}